\documentclass[twocolumn]{aastex631}

\newcommand\chandra{{\it Chandra}}
\begin{document}

\title{Probing AGN-ISM Feedback through Extended X-ray Emission in ESO 137-G034}
\author[0000-0002-3626-5831]{D.~{\L}.~Kr\'{o}l}
\affiliation{Center for Astrophysics $|$  Harvard \& Smithsonian, 60 Garden Street, Cambridge, MA 02138, USA}
\affiliation{Astronomical Observatory of the Jagiellonian University, Orla 171, 30-244 Krak\'{o}w, Poland}
\author[0000-0002-3554-3318]{G.~Fabbiano}
\affiliation{Center for Astrophysics $|$  Harvard \& Smithsonian, 60 Garden Street, Cambridge, MA 02138, USA}
\author[0000-0001-5060-1398]{M.~Elvis}
\affiliation{Center for Astrophysics $|$  Harvard \& Smithsonian, 60 Garden Street, Cambridge, MA 02138, USA}
\author[0000-0001-8112-3464]{A.~Trindade~Falcão}
\affiliation{Center for Astrophysics $|$  Harvard \& Smithsonian, 60 Garden Street, Cambridge, MA 02138, USA}
\affiliation{ NASA-Goddard Space Flight Center, Code 662, Greenbelt, MD 20771, USA}
\author[0000-0001-9815-9092]{R.~Middei}
\affiliation{Center for Astrophysics $|$  Harvard \& Smithsonian, 60 Garden Street, Cambridge, MA 02138, USA}
\affiliation{INAF Osservatorio Astronomico di Roma, 10
Via Frascati 33, 00078 Monte Porzio Catone, RM, Italy}
\affiliation{Space Science Data Center, 12 Agenzia Spaziale Italiana, Via del Politecnico snc, 00133 Roma, Italy}

\author[0000-0002-0001-3587]{D.~Rosario}
\affiliation{School of Mathematics, Statistics and Physics, Newcastle University, Newcastle upon Tyne NE1 7RU, UK}
\author[0000-0003-4949-7217]{R.~Davies}
\affiliation{Max Planck Institute for Extraterrestrial Physics, Giessenbachstrasse, 85741 Garching bei München, Germany}
\author[0000-0002-2125-4670]{T.~Shimizu}
\affiliation{Max Planck Institute for Extraterrestrial Physics, Giessenbachstrasse, 85741 Garching bei München, Germany}
\author{D.~Hill}
\affiliation{School of Mathematics, Statistics and Physics, Newcastle University, Newcastle upon Tyne NE1 7RU, UK}

\begin{abstract}
We present a detailed analysis of the X-ray emission of the Compton-thick (CT) AGN ESO\,137-G034 based on deep ($\sim230$\,ks) Chandra observations. As in other CT AGNs, the morphology of the emission is elongated, approximately following the [O III] ionization bicone. With spatially resolved spectral modeling, we show that the extended emission within the bi-cone regions { is most readily explained} as from a mixture of photo-ionized gas and shock-heated plasma, reflecting the combined effects of radiative and kinematic AGN feedback. By comparing the morphology of the { X-ray emission in narrow spectral bands and that of the 3\,cm radio jet, we find suggestive evidence of thermal, possibly shocked emission associated with the SE termination of the radio jet. This interpretation is also supported by the lack of [O~III]   relative to the $0.3-3.0$ keV flux in the inner $3^{\prime\prime}$ ($\sim600$\,pc) of the SE cone, which would be consistent with an additional thermal X-ray component on top of the photonized emission of an outflowing wind. A similar effect is only seen within the inner $1^{\prime\prime}$ ($200$\,pc) of the NW cone. In the radial profile of the [OIII]/X-ray flux ratio and the X-ray hardness ratio within the inner $\sim3^{\prime\prime}$ ($\sim600$\,pc) of the SE cone we see an asymmetry, with no counterpart in the NW cone. We detect soft extended X-ray emission in the cross-cones which may originate from the interaction of an embedded radio jet with a clumpy interstellar medium (ISM). }These results highlight the { importance of both} radiative and mechanical feedback in shaping the circumnuclear environment of ESO\,137-G034.
\end{abstract}


\section{Introduction} \label{sec:intro}
The disagreement between galaxies' observed luminosity function and the mass function predicted by the $\Lambda$CDM simulations of the large scale structures is a long-standing problem, since the observed distribution deviates from the shape predicted by simulations, especially for the lowest and highest galactic masses \citep{Benson03,Schechter76}. The observed lower number of low-mass galaxies can be tied to the impact of supernovae on the interstellar medium (ISM) \citep{Keres09}, but to explain the deficit of the highest luminosity galaxies, a much more powerful mechanism is required. Since most massive galaxies are believed to harbor a supermassive black hole (SMBH), active galactic nuclei (AGN) are a compelling source of the required energy \citep{Fabian12}. 
Strong coupling of AGN and galaxy evolution is also supported by a tight correlation between the SMBH and host galaxy properties \citep[e.g., $M-\sigma$ relation][]{Kormendy13}, making the understanding of AGN-ISM interactions, known as AGN feedback, an important piece of understanding galaxy evolution. 
This picture becomes more complex as new results obtained by JWST suggest that the SMBH-to-galaxy mass ratios at high redshifts are extremely high, indicating either very rapid initial BH growth and/or a heavy seed origin \citep[see, e.g.,][]{Adamo24, Bogdan24, Maiolino24}. 

While population studies can provide insights into the role of AGN in galaxy evolution, high-resolution observations of AGN offer the unique opportunity to observe feedback as it occurs, allowing to probe different feedback avenues. In particular, high-resolution X-ray observations of the extended emission around Compton Thick (CT) AGN can shed light on the complex nature of AGN-ISM interactions, as X-rays are particularly effective in tracing the mechanisms behind this coupling, including ISM photoionization, jet-ISM shocks, and large-scale winds. Thanks to the CT nature of these sources, the central continuum exhibits a lower apparent flux, enabling detailed analysis of the diffuse emission \citep[see recent review by][]{Fabbiano22rev}.

\begin{figure}
    \centering
    \includegraphics[width=0.5\textwidth]{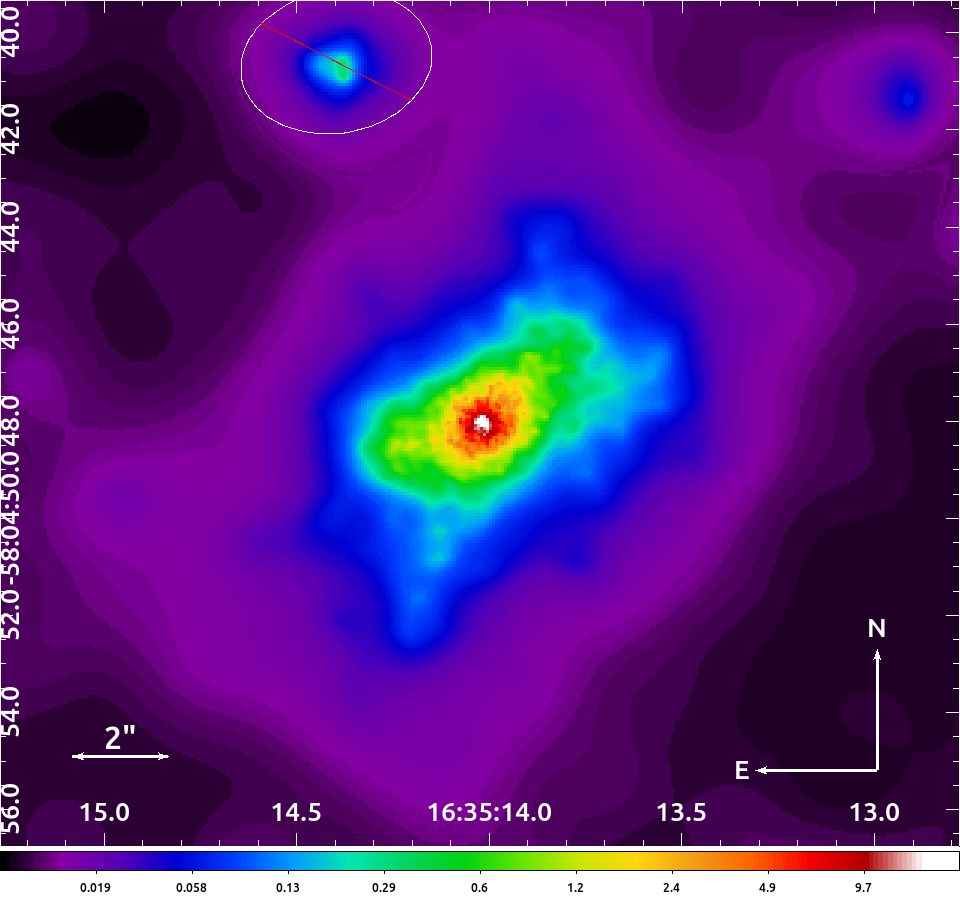}
    \caption{{\it Chandra} {ACIS-S3 (CCD ID=7)} image of ESO\,137-G034, merged from observations listed in Tab.\,\ref{tab:observations}. The image has been binned with 1/8 pixel size and adaptively smoothed. Crossed circle marks the point source excluded from the analysis.}
    \label{fig:merged}
\end{figure}

The diffuse X-ray emission of Compton-Thick (CT) AGN tends to roughly trace the bi-cone morphology of regions characterized by high [OIII]/H${\alpha}$ ratios in optical data \citep{Schmitt07,Fischer13}. These ionization cones are believed to be excited by AGN nuclear radiation ``collimated'' by the torus in the standard AGN model \citep{Wilson96}. X-ray data with sufficient spatial and spectral resolution add much more to the story. Spatially resolved X-ray spectroscopy reveals that the emission from cone regions cannot be characterized purely by photoionized plasma emission, and requires an additional thermal plasma component \citep[e.g.,][]{Fabbiano18b,Wang11a,Travascio21,Trindade23}. Shock-heated thermal plasma can be traced by an excess { of the emission at $\sim1$\,keV, an energy band containing the Ne~IX and Ne~X lines, as well as  a contribution from the broader  Fe-L complex} \citep[see][]{Wang11a,Paggi12}, which serves as a tracer of this thermal component. Moreover, diffuse X-ray emission extends in both the bi-cone and cross-cone directions \citep{Fabbiano18b,Ma20}, hinting at either a ``leaky'', porous torus geometry \citep{Fabbiano18b} or interaction of a radio jet with a patchy, dense ISM \citep{Fabbiano22}.

In this paper, we analyze deep {\it Chandra} ($\sim235$\,ks) observations of ESO\,137-G034, with the goal of disentangling different emission mechanisms, and corresponding feedback modes, in the cone and cross-cone regions. ESO\,137-G034 is an S0/a galaxy hosting a Seyfert type II AGN \citep{Ferruit00} at $z=0.009$ \citep[$D_{L}=41.21$\,Mpc; $1^{\prime\prime}\simeq200\textrm{pc}$][]{Koss22}. ESO\,137-G034 was part of a program targeting AGN in search of extended hard X-ray emission \citep{Ma20}, showing a clear extent in $50$\,ks {\chandra}    observations.

In Sec.\,\ref{sec:data_analysis}, we describe the observations, astrometric alignment, and data reduction. In Sec.\,\ref{sec:morphology}, we analyze the morphology of the extended emission through azimuthal and radial profile extractions across different energy bands and regions. In Sec.\,\ref{sec:spectra}, we model the spectra of the various source regions. In Sec.\,\ref{sec:discussion}, we interpret our results and discuss the origin of the emission in different parts of the source. In Sec.\,\ref{sec:summary}, we summarize our findings.

In all calculations we assumed flat $\Lambda$CDM cosmology with $H_0=69.3$\,km\,s$^{-1}$\,Mpc$^{-1}$ and $\Omega_m=0.287$ \citep{Hinshaw13}. Though the paper we refer to the soft energy range as  $0.3-1.5$\,keV, medium as $1.5-4.0$\,keV and hard as $4.0-7.0$\,keV. All position angles are measured from west to north.
\section{Data Reduction}\label{sec:data_analysis}

\begin{deluxetable}{ccccc}[t]
\label{tab:observations}
\tablecaption{Details on Analyzed Data, including the ObsID number, date of the observation, exposure time and astrometry offset ($\Delta$x and $\Delta$ y), see Sec.\,\ref{sec:data_analysis} for the details. }
\tablehead{\colhead{ObsID } & \colhead{Date} & \colhead{Exposure Time} & \colhead{$\Delta$x} &\colhead{$\Delta$y}  \\ & \colhead{} & \colhead{[ks]} & \colhead{pixels} & \colhead{pixels}}
\startdata
    21422& 2020-10-18  & 44.76  &   reference \\
    25268&  2023-03-29  & 10.13  &  $-1.95 $ &   $-1.53 $  \\
    25842& 2023-10-28  & 9.64 	  &  $-0.85 $ & $-0.52$  \\
    25843&  2023-03-10 &  29.15 &  $-1.51 $  &   $-0.78 $ \\
    25844& 2022-01-12  &  22.41 &  $-1.17 $   & $-2.09 $\\
    25845& 2023-06-28  & 28.50  &  $1.34$  &   $-0.92 $     \\
    25846& 2022-01-14  & 10.52 	    &  $0.43 $ &   $-1.59 $    \\
    27770&   2023-03-30&  34.05 	 &  $-1.70 $&      $-0.30 $  \\
    28912&  2023-09-15 &  26.39 &  $-0.17 $&  $-1.17 $    \\
    28995&   2023-10-28& 	9.66  &  $0.55 $&  $-0.39 $    \\
    28996& 2023-10-29   &  8.99  &  $0.70 $  &   $0.38 $   \\
\hline
\enddata
\tablecomments{Data set DOI: \url{https://doi.org/10.25574/cdc.374}}
\end{deluxetable}

\begin{figure*}[thp!]
    \centering
    \includegraphics[width=0.56\textwidth]{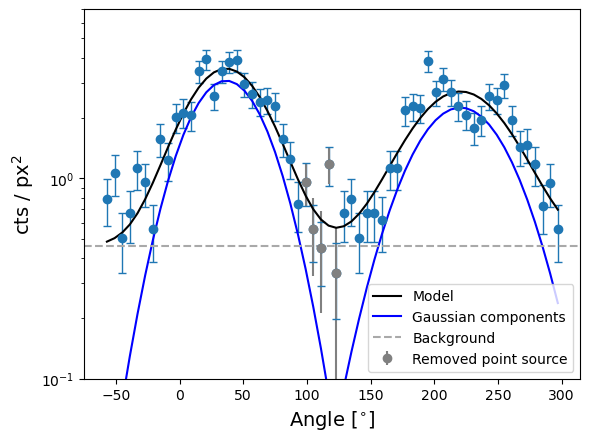}
    \includegraphics[width=0.42\textwidth]{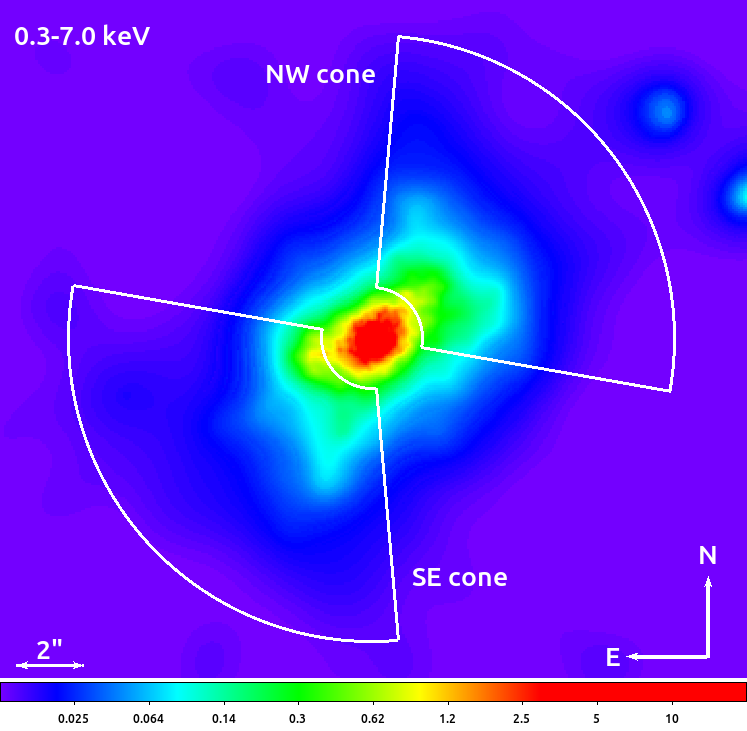}
    \includegraphics[width=0.56\textwidth]{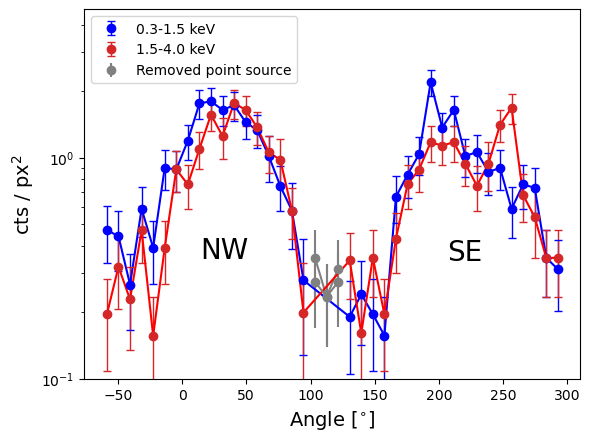}
    \includegraphics[width=0.42\textwidth]{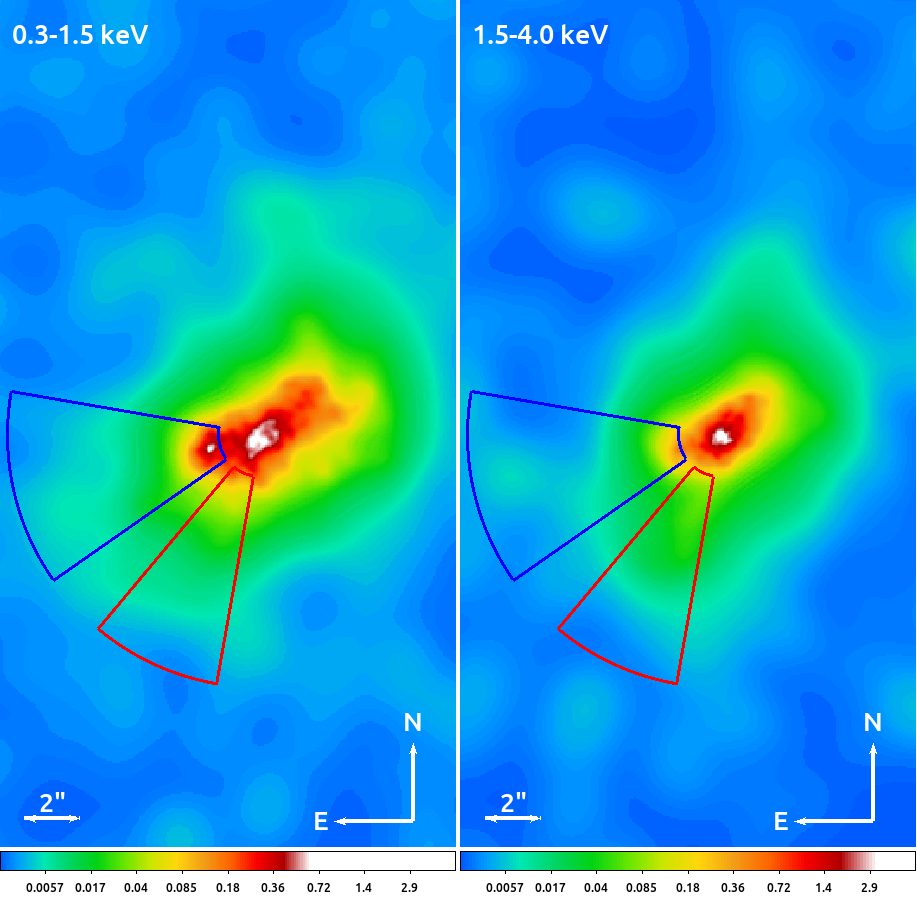}
 \caption{Left panels: Azimuthal profiles of the extended emission extracted between $1.5^{\prime\prime}$ and $9^{\prime\prime}$ from the nucleus, based on the merged \chandra\ data image. The upper left panel shows the profile for the $0.3$--$7.0$\,keV energy range, overlaid with the best-fit model (dark gray line), which consists of two Gaussian components (dark blue) and a constant background (dashed gray). Gray points indicate directions where point sources were removed. The lower left panel shows azimuthal profiles for two energy ranges: $0.3$--$1.5$\,keV (blue line) and $1.5$--$4.0$\,keV (red line). Position angles are measured from the West to the North on the sky. Right panels: The upper right panel displays the adaptively smoothed, merged image in the $0.3$--$7.0$\,keV range, overlaid with the azimuthal profile extraction region. Directions of peak emission in the $0.3$--$1.5$\,keV and $1.5$--$4.0$\,keV bands are marked in blue and red, respectively. The lower right panel shows the adaptively smoothed emission in the $0.3$--$1.5$\,keV and $1.5$--$4.0$\,keV bands (from left to right), Blue and red cones mark angular position of peaks in $0.3$--$1.5$\,keV and $1.5$--$4.0$\,keV respectively.  }
    \label{fig:azimuthal}
\end{figure*}
ESO\,137-G034 was observed 11 times  (see exposures details in Tab.\ref{tab:observations}) by \chandra's \citep{Weisskopf02}  Advanced CCD Imaging Spectrometer (ACIS) \citep{Garmire03}. We combined all available data and followed standard \chandra\ data reduction procedures using  {\tt CIAO v4.16} \citep{Fruscione06} {with {\tt CALDB v4.11.5}}, starting with the {\tt chandra\_repro}\footnote{Description of all CIAO scripts can be found at \url{https://cxc.cfa.harvard.edu/ciao/ahelp/index_alphabet.html}} reprocessing script. To ensure that astrometry corrections are calculated with respect to position of the nucleus, we followed the approach by \cite{Fabbiano17}. Offsets were calculated based on the position of a centroid around the maximum counts pixel in the $5-7$ keV energy range for each obsID. We chose the longest observation (21422) as the reference. Analysis was carried out on images binned with $1/8$ of ACIS native pixel size and then smoothed with 2\,pixel Gaussians. The derived offsets are listed in Tab.\,\ref{tab:observations}. The resulting adaptively smoothed merged image in the $0.3-7.0$\,keV band is shown in Fig.\,\ref{fig:merged}. The X-ray source is clearly extended to the NW and SE on scales of {$\sim10^{\prime\prime}$} ($2$\,kpc) from the nucleus, in what may be a bicone structure as seen in many other CT AGNs \citep[][]{Fabbiano22rev}.
Since we are interested in the properties of any extended emission, to identify point sources { so that they could be excluded from the analysis,} we employed the {\tt wavdetect} algorithm {with the { \tt scales} parameter set to: $1$\,px, $2$\,px, $4$\,px, $8$\,px, and $16$\,px. Other parameters of the script were set to their default values. Only one point source was detected within $9^{\prime\prime}$ of the nucleus. We removed it and replaced it with pixel values interpolated from the data at the same distance from the nucleus using the {\tt dmfilth} procedure. The area corresponding to the removed point was excluded from spectral analysis and is marked in gray in all radial and azimuthal profiles}.

\begin{deluxetable}{ccccc}[t]
\label{tab:excess}
\tablecaption{Background Subtracted Excess Data Counts over the Model PSF for $1.5^{\prime\prime}-8^{\prime\prime}$ Conical Regions }
\tablehead{\colhead{Energy } & \colhead{Cone NW} & \colhead{Cone SE} & \colhead{Cross-cone SW} & \colhead{Cross-cone NE}\\ \colhead{keV} & \colhead{cts/px$^2$} & \colhead{cts/px$^2$} & \colhead{cts/px$^2$} &\colhead{cts/px$^2$}}
\startdata
$ 0.5-7.0 $ & $683 \pm 28 $ &  $647 \pm 28 $ &  $134 \pm 15 $ &  $118 \pm 15 $\\
$ 0.5-1.5 $ & $333 \pm 19 $ &  $314 \pm 18 $ &  $82 \pm 10 $ &  $57 \pm 8 $\\
$ 1.5-3.0 $ & $256 \pm 17 $ &  $230 \pm 16 $ &  $39 \pm 8 $ &  $34 \pm 8 $\\
$ 3.0-4.0 $ & $40 \pm 8 $ &  $44 \pm 8 $ &  $4 \pm 4 $ &  $4 \pm 4 $\\
$ 4.0-5.0 $ & $14 \pm 6 $ &  $20 \pm 6 $ &  $-1 \pm 4 $ &  $7 \pm 5 $\\
$ 5.0-6.0 $ & $5 \pm 5 $ &  $9 \pm 5 $ &  $3 \pm 4 $ &  $-3 \pm 4 $\\
$ 6.0-7.0 $ & $23 \pm 8 $ &  $15 \pm 7 $ &  $-2 \pm 5 $ &  $5 \pm 6 $\\
\hline
\enddata
\end{deluxetable}

\section{Morphology of X-ray Extended Emission }\label{sec:morphology}

\subsection{Azimuthal profiles}\label{sec:azimuthal}

To determine the core and cross-cone extraction regions, we examined the azimuthal profile of the extended emission. In particular, we extracted profiles binned in $6^\circ$ intervals within an annulus spanning $1.5^{\prime\prime}$ to $9^{\prime\prime}$ from the nucleus (see top-left panel of Fig.\,\ref{fig:azimuthal}). The profile was fitted with a combination of two Gaussians of the form $A_i \exp\left[-\frac{(\phi - \phi_i)^2}{2 \sigma_i^2}\right]$ plus a constant background. The best-fit parameters are: $\phi_1=(35\pm 3)^\circ$, $\phi_2=(219 \pm 3)^\circ$, $\sigma_1=(38\pm4)^\circ$, $\sigma_2=(30\pm3)^\circ$, $A_1=(2.7\pm0.2)$\,cts/px$^2$, and $A_2=(3.6\pm0.3)$\,cts/px$^2$. 
Based on that  we estimate the bicone region to be comprised between $-10^\circ$--$85^\circ$ (NW) and $170^\circ$--$275^\circ$ (SE). We define  cross-cone the complementary regions. These regions are indicated in the top right panel of Fig.\,\ref{fig:azimuthal}.

While the overall X-ray extended emission morphology follows the bi-cone/cross-cone structure, but the azimuthal profile of the south-east (SE) cone shows two distinct maxima at $\sim195^{\circ}$ and $\sim255^{\circ}$ as indicated in the bottom-left panel of Fig.\,\ref{fig:azimuthal}. These peaks are pronounced in different energy bands:  $\sim195^{\circ}$ maxima more at the soft ($0.3$--$1.5$\,keV) energy band, while the $\sim255^{\circ}$ peak appears in medium ($1.5$--$4.0$\,keV) energy band (see the profiles and images in the bottom panels of Fig.\,\ref{fig:azimuthal})

\begin{figure*}
    \centering
    \includegraphics[width=1.0\textwidth]{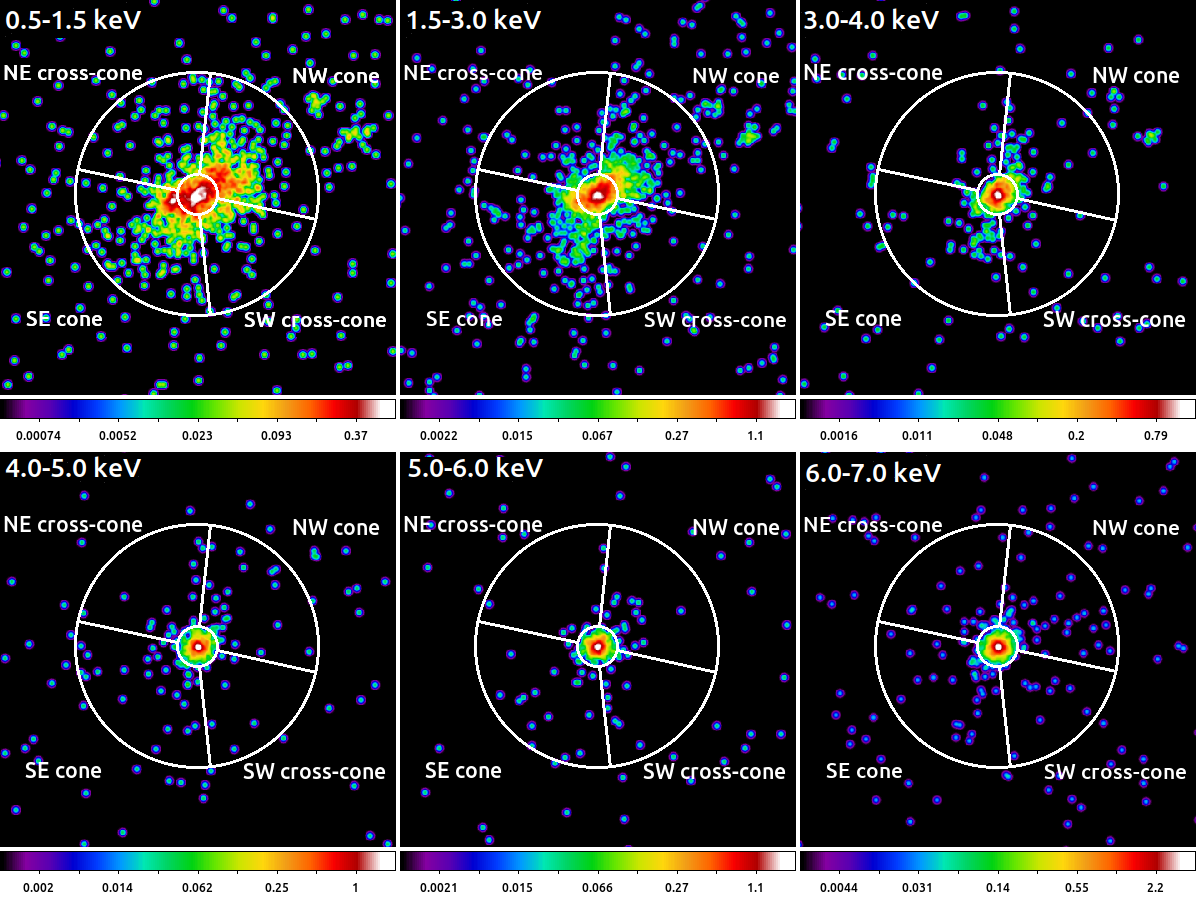}
\caption{Images of ESO\,137--G034 in different energy bands{, starting from the left-top corner: $0.5$--$1.5$\,keV, $1.5$--$3.0$\,keV, $3.0$--$4.0$\,keV, $4.0$--$5.0$\,keV, $5.0$--$6.0$\,keV, and $6.0$--$7.0$\,keV. Contours mark } the spectral extraction regions. The images were binned to 1/8\,px resolution and smoothed using a Gaussian kernel with $\sigma=2$\,px.}
    \label{fig:regions_radial}
\end{figure*}

\begin{figure*}[h!]
    \centering  
    \includegraphics[width=0.9\textwidth]{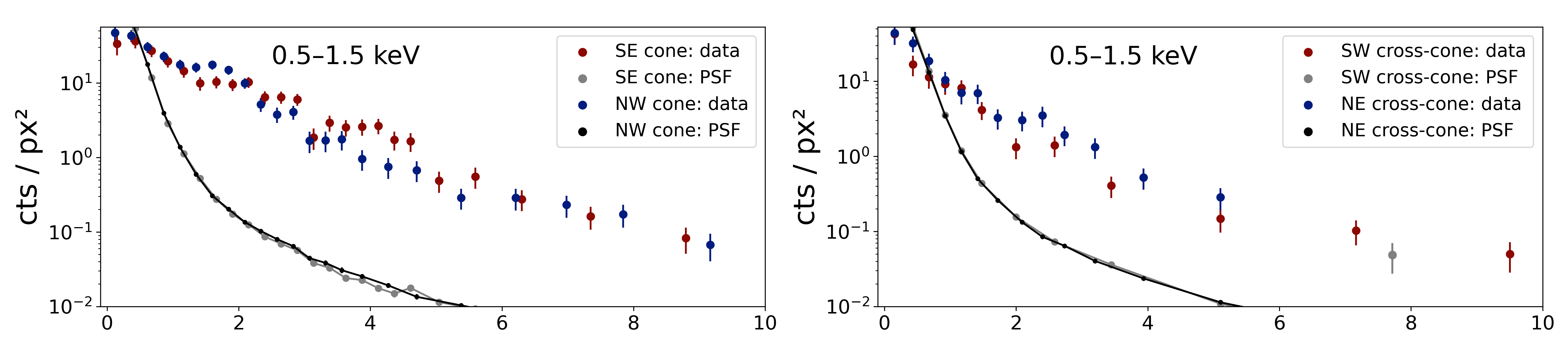}
    \includegraphics[width=0.9\textwidth]{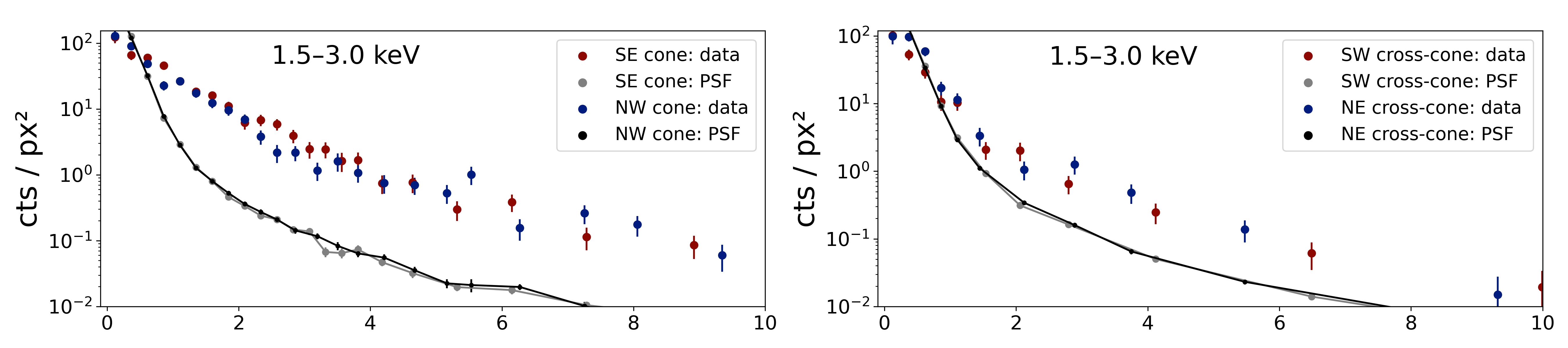}
    \includegraphics[width=0.9\textwidth]{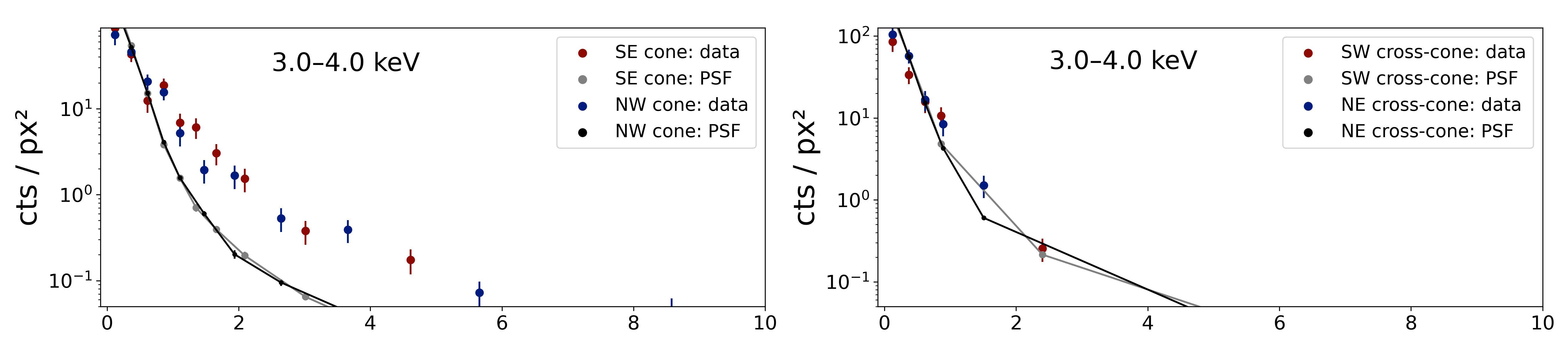}
    \includegraphics[width=0.9\textwidth]{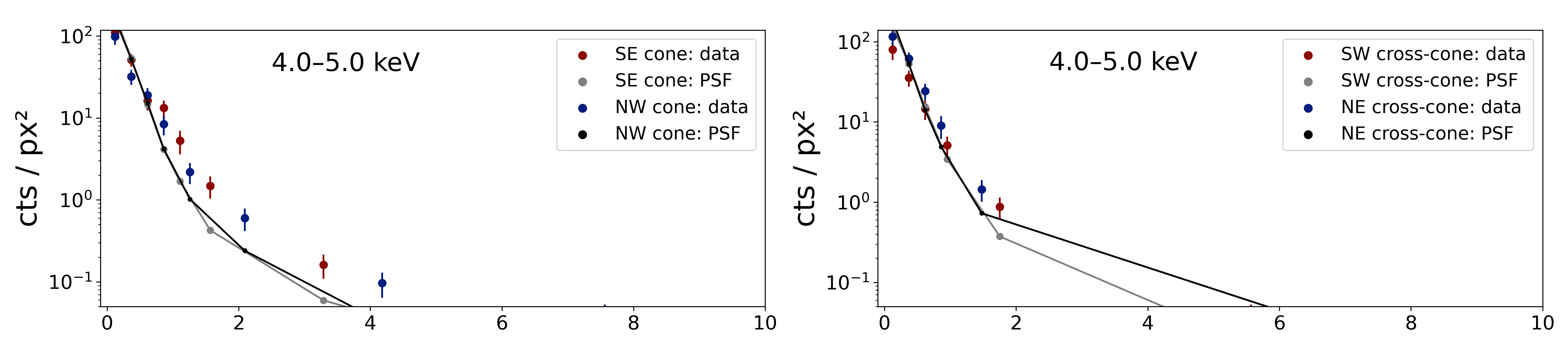}
    \includegraphics[width=0.9\textwidth]{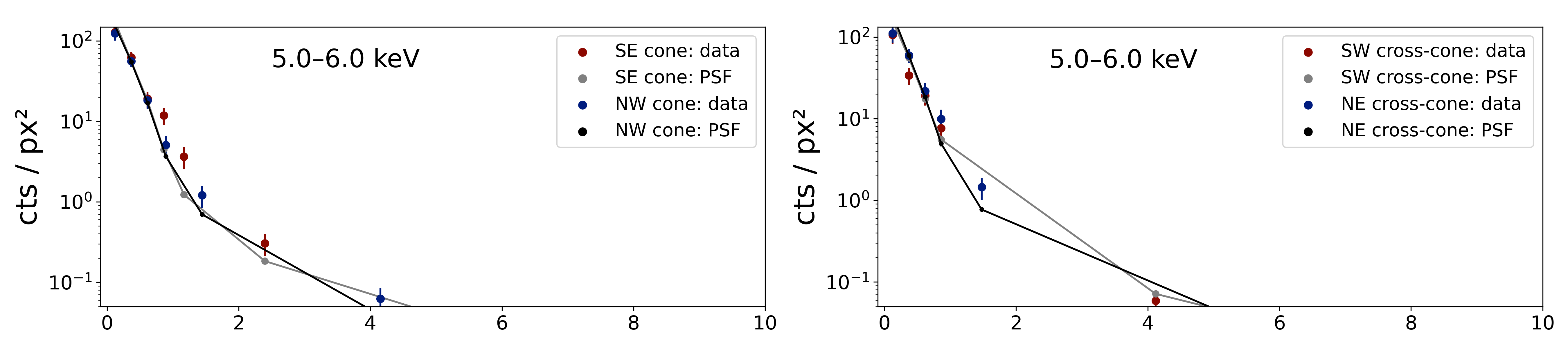}
    \includegraphics[width=0.9\textwidth]{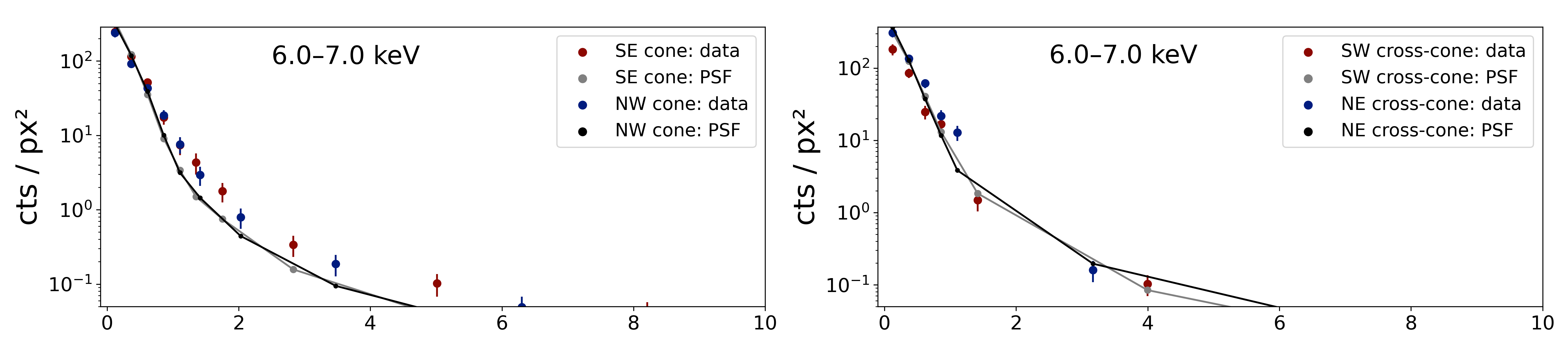}
    \caption{Background-subtracted surface brightness radial profiles in the cone regions (left column) and cross-cone regions (right column) with corresponding PSF profiles, for different energy ranges, from the top to the botom: $0.5$--$1.5$\,keV, $1.5$--$3.0$\,keV, $3.0$--$4.0$\,keV, $4.0$--$5.0$\,keV, $5.0$--$6.0$\,keV, and $6.0$--$7.0$\,keV. {Error bars correspond to $1\sigma$ confidence intervals.} }
    \label{fig:radial_1}
\end{figure*}

\subsection{Radial profiles}

\subsubsection{PSF simulations}\label{subsec:psf}
To assess the significance of the extended emission in different energy ranges, we performed simulations of the \chandra\ point spread function (PSF). First, for each obsID, we extracted the nuclear spectra from a circular region of radius $1.5^{\prime\prime}$ using the {\tt specextract} routine and fitted phenomenological models to describe the spectral shape. 

Next, we simulated the \chandra\ PSF using {\tt ChaRT} \citep{Carter03} and {\tt MARX} \citep{Davis12}, with the fitted spectral model as input. For each pointing, we performed 50 PSF simulations using {\tt ChaRT}, { and derived the mean PSF, while the differences in the PSF realizations were adopted as errors.} Each resulting ray file was then converted to a pseudo-event file with {\tt MARX}, assuming an ASPECT blur of $0.07$ \citep[see Appendix 5 in][]{Ma23}. Increasing the ASPECT blur to $0.20$ does not significantly change the results. All simulations were then combined into a single pseudo-event file using the {\tt dmmerge} procedure. { Finally, we normalized the PSF to match the counts number in inner $1.5^{\prime\prime}$ of the source in each energy band. }

\subsubsection{Radial Emission Profiles}
\label{sec:radial}

The cone and cross-cone regions are overlaid on the emission maps in different energy bands in Fig.\,\ref{fig:regions_radial}. Using the merged data with 1/8 px binning, we extracted radial surface brightness profiles for each region in six energy bands: $0.5$--$1.5$\,keV, $1.5$--$3.0$\,keV, $3.0$--$4.0$\,keV, $4.0$--$5.0$\,keV, $5.0$--$6.0$\,keV, and $6.0$--$7.0$\,keV. The minimum radial bin size was set to $0.246^{\prime\prime}$ and adjusted to ensure at least 10 counts per bin. 

The same regions were used to extract the PSF radial profiles, which were normalized to match the number of counts in the { inner $1.5^{\prime\prime}$} central region. In Fig.\,\ref{fig:radial_1}, we present the background-subtracted radial surface brightness profiles for both the data and PSF. Table\,\ref{tab:excess} shows the excess of counts in the data over the PSF for the cone and cross-cone regions across energy bands.
{ From $0.5$ to $3$\,keV, }both cone regions show { highly significant extended surface brightness ($>14\sigma$). In the $3.0-4.0$\,keV range, the surface brightness is extended at $>5\,\sigma$ significance. In $4$--$5$\,keV energy range, only the SE cone shows extended emission with a significance $>3\sigma$.} There is no significant extension in the $5$--$6$\,keV band.  In the $6$--$7$\,keV range, the presence of extended emission is confirmed at the $\sim3\sigma$ level in the NW cone and at the $\sim2\sigma$ level in the SE cone. The cross-cone regions show extended emission with $>5\sigma$ significance up to $3$\,keV, after which the radial profiles become consistent with the nuclear PSF.

\section{Spectral Analysis}
\label{sec:spectra}
\subsection{Extraction of Spectra}

Source and background energy spectra were extracted separately for each observation. We used a circular region for the nucleus ($r_N = 1.5^{\prime \prime}$) and {\tt panda} regions with inner radius $r_{\rm in}=1.5^{\prime \prime}$ and outer radius $r_{{out}}=9^{\prime \prime}$ for the cone and cross-cone regions, using the same position angle definitions as in Sec.\,\ref{sec:radial}. Point sources identified with the {\tt wavdetect} tool were excluded. For the background, we used a concentric {\tt panda} region with $r_{\rm in} = 10^{\prime \prime}$ and $r_{\rm out}=12^{\prime \prime}$, excluding two point sources at that radius. The spectra were combined using the {\tt combine\_spectra} script, and in all cases, the background was subtracted before modeling.

Initially, spectra from the two cone regions were extracted and fitted independently. As the best-fit model parameters were consistent within uncertainties, we combined the spectra from both cones to improve data statistics. The same procedure was followed for the cross-cone regions, where photon statistics are even more limited.

\subsection{Spectral Models}\label{sec:models_components}

{Spectral fitting was performed with {\tt Sherpa v4.16.0} \citep{Siemiginowska24}. 
{ Fit quality was assessed based on the {\tt wstat} statistic}\footnote{see  \url{https://sherpa.readthedocs.io/en/latest/statistics/api/sherpa.stats.WStat.html}} \citep{Cash79} { and the presence of significant localized residuals \citep{Fabbiano18b}.  Data were binned with a minimum of $10$ counts per bin. { The fitting results are the same for $\chi^2$ statistic, with variance calculated from the data}\footnote{{\tt chi2datavar}, see  \url{https://cxc.cfa.harvard.edu/sherpa/statistics/index.html\#chidata}}. Fits were performed in the $0.3$--$7.0$\,keV energy range. All provided errors represent $1\sigma$ confidence intervals, calculated with the {\tt conf()} function implemented in {\tt sherpa}. Unless specified otherwise, results and definitions are provided in cgs units.}
}

To model the photoionized spectra, we constructed a grid of models using {\tt Cloudy c.23.01} \citep{Gunasekera23}. Following \cite{Kraemer20}, we assumed an incident radiation field in the form of a broken power-law, $L \propto \nu^{\alpha_i}$, with $\alpha_1=1.0$ for $h\nu \in (10^{-4},13.6)$\,eV, $\alpha_2=1.3$ for $h\nu \in (13.6,500)$\,eV, and $\alpha_3=0.5$ for $h\nu \in (0.5,300)$\,keV and {$1.5\times$ solar abundance}. We assumed a turbulent velocity of $100$\,km s$^{-1}$ and a hydrogen density of $10^{4.5}$\,cm$^{-3}$. The grid spanned the range $\log U = [-2, 2]$ and $\log N_H = [21, 25]$ with a step size of 0.1. {$U$ is the dimensionless ionization parameter defined as $\frac{\Phi(H)}{cn(H)}$, the ratio of ionizing photons flux $\Phi(H)$, and the the total hydrogen density n$(H)$, with $c$ marking the speed of light. $\log N_H$ is  a common logarithm of the hydrogen column density in atoms per cm$^{2}$ units. } 

{The spectral models also included {\tt xsapec} and {\tt xszphabs}, which are {\tt xspec} \citep{Arnaud99} implementations of the Astrophysical Plasma Emission Code \citep[APEC, ][]{Smith01} and photoelectric absorption \citep{Balucinska92} models. {\tt xsapec} parameters include plasma temperature, $kT$, abundance, redshift, and normalization, {\tt norm}, in norm$_{a} = \frac{10^{-14}}{4\pi D_A^2(1+z)^2}\int \textrm{d}V\, n_e n_H$ units. $D_A$ denotes the angular diameter distance to the source, $n_e$ and $n_H$ are the electron and ion densities, and d$V$ is the volume element. Two other model parameters-redshift, $z$, and metallicity-were frozen at the source redshift and solar metallicity, respectively. {\tt xszphabs} is described by the equivalent hydrogen column density, $ N_H$, and redshift, $z$.}

{The {\tt mytorus} model  was employed to describe the torus-reprocessed nuclear emission \citep{Yaqoob12}. We included Compton-scattered power-law continuum and fluorescent Fe K{$\alpha$} line emission components. }Model parameters include intrinsic continuum photon index, $\Gamma$, component normalization, norm, defined as photon flux at $1$\,keV,  { equatorial} column density $N_H$, inclination angle  between the torus polar axis and the observer's line of sight in degrees,  and redshift. 

For the soft X-ray emission, we tested several model combinations: a single {\tt xsapec}, a single {\tt cloudy}, a combination of both, and models with two {\tt xsapec} or two {\tt cloudy} components. Galactic absorption was { always } included via the {\tt xsphabs} component, with a fixed hydrogen column density of $N_H = 2.36 \times 10^{20}$\,cm$^{-2}$ \citep{HI4PI16}. { Presence of a second  {\tt xsphabs} component was tested to account for the intrinsic absorption.}

\subsection{Fitting Results}

\begin{figure*}[thp!]
    \centering
    \includegraphics[width=0.45\textwidth]{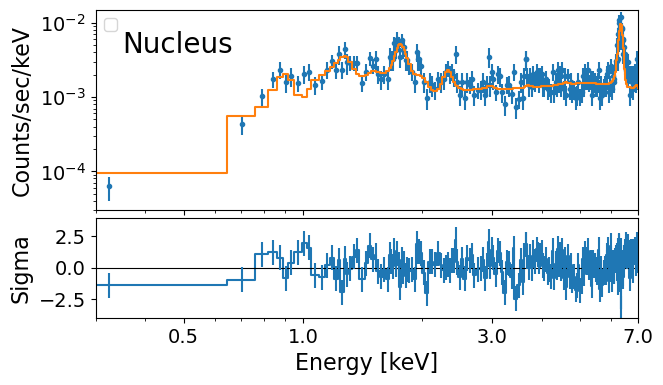}
    \includegraphics[width=0.45\textwidth]{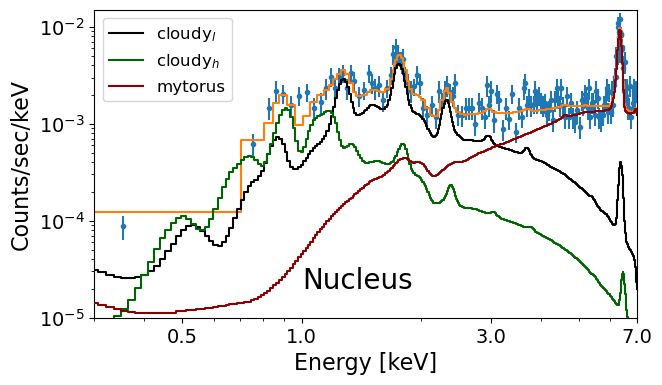}
    \includegraphics[width=0.45\textwidth]{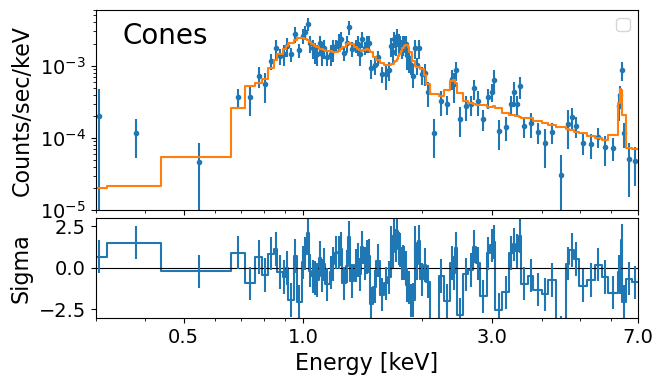}
    \includegraphics[width=0.45\textwidth]{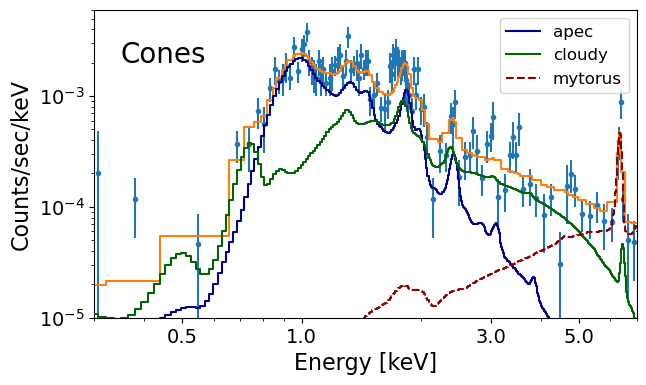}
    \includegraphics[width=0.45\textwidth]{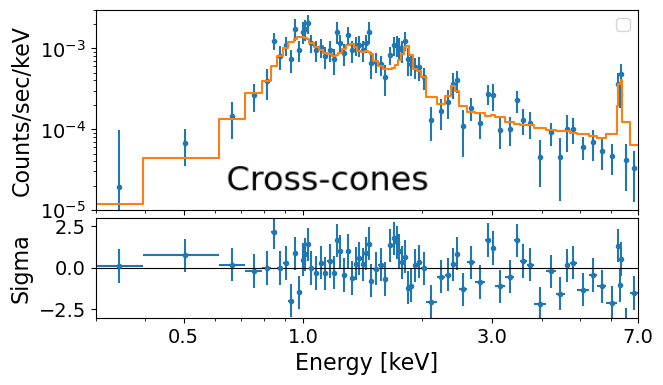}
    \includegraphics[width=0.45\textwidth]{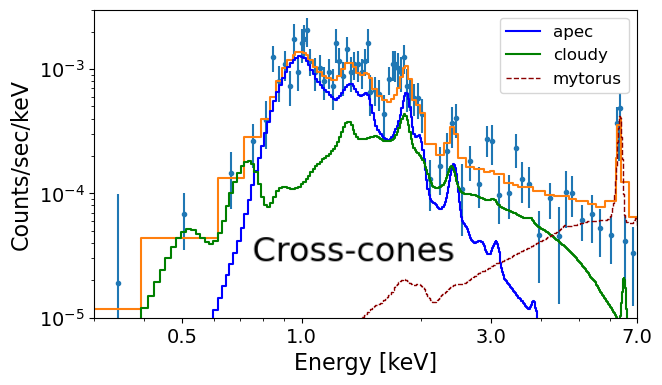}
    \includegraphics[width=0.45\textwidth]{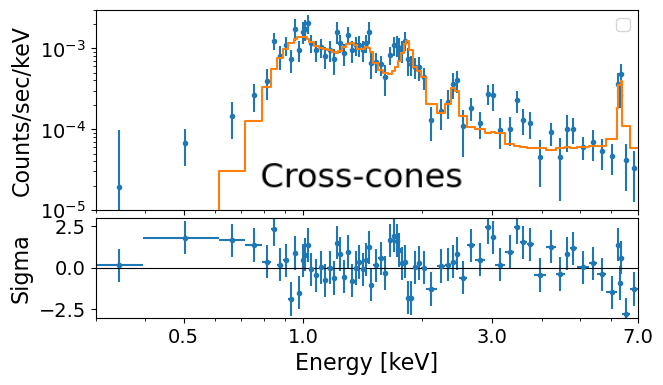}
    \includegraphics[width=0.45\textwidth]{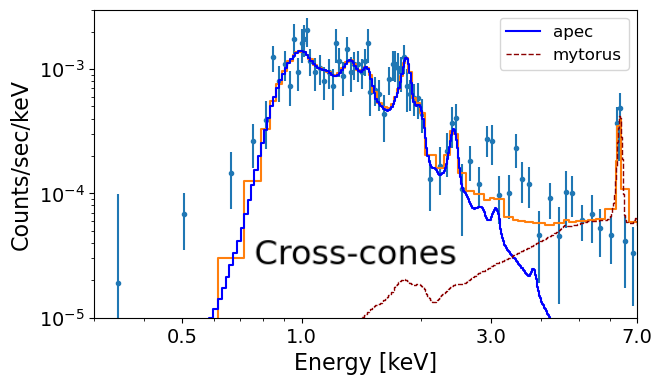}
    \caption{Spectrum of the nucleus ({ first} row), bi--cone region ({ second} row), and cross-cones (two bottom rows). The left column shows the best-fit model and its residuals, while the right column presents the components of the best-fit models.}
    \label{fig:spectra}
\end{figure*}

\subsubsection{Nucleus}\label{sec:nucleus_fit}

The best-fit model for the nuclear spectrum consists of two {\tt cloudy} components with an internal absorption, {\tt xsphabs},  and a {\tt mytorus} model. {Single {\tt apec} and single {\tt cloudy} models do not provide satisfactory fit to the data (see Appendix\,\ref{appendix:spectra} for the f-test results).  
The soft part of the spectra can also be described by a combination of {\tt apec} and {\tt cloudy},  with slightly worse fit statistics ({ {\tt wstat} / d.o.f.$=1.07$ and $1.18$, }respectively) and {\tt apec} best-fit temperature $kT\sim0.1$\,keV.  This model can be however rejected based on the unrealistically high intrinsic luminosity of the {\tt apec} component ($\sim 10^{44}$erg s$^{-1}$ in $0.05-4$\,keV energy range). A combination of two thermal components leads to poorer fit quality, with { {\tt wstat} / d.o.f. } $=1.24$. During the fitting procedure,} we fixed the inclination angle in the {\tt mytorus} model to $80^\circ$, as appropriate for a CT AGN in which the torus is blocking our view of the central AGN, and set the {intrinsic power-law continuum photon index for}  both the continuum and line components to $\Gamma=1.8$, a typical value for AGN \citep[e.g.,][]{Marchesi17}.{The equatorial column densities ($N_{HFe}$ and $N_{Hc}$), and the normalizations (norm$_{Fe}$ and norm$_{ c}$) were free model parameters. Fe and c indices denote the fluorescent Fe line and continuum components respectively.}


The fitted parameters of  the {\tt cloudy} models include  ionization parameter ($\log U$) the gas column density ($\log N_H$) and the normalization. The soft emission ($<4.0$\,keV) is dominated by a combination of a highly ionized, low-column-density component ($\log U_h \sim 0.6$, $\log N_{Hh} \sim 19$) that dominates below $1$\,keV, and a low-ionization, high-column-density component ($\log U_l \sim -1.4$, $\log N_{Hl} \sim 22$), that dominates in the feature-rich $1-3$\,keV band. { As the best-fit of the intrinsic absorption gives only an upper limit on the equivalent column density ($N_H <3\times10^{20}$atoms cm$^{-2}$), we removed it from the final model.  We denote ionized, low-column-density  {\tt cloudy} component with $h$ and  low-ionization, high-column-density  {\tt cloudy} component with $l$.} 

The hard X-ray emission ($>4$\,keV) is dominated by the Compton-scattered component from {\tt mytorus}. The iron line and { Compton-scattered }continuum emission are absorbed by column densities that are compatible within errors, { therefore $ N_{Hc}$ and $N_{HFe}$ parameters were linked, and their best-fit value is reported in the Tab.~\ref{tab:spectral_fit} as $N_H$}. { An additional absorbed power-law model component, which would account for the transmitted continuum, does not improve the fit statistics}.

\begin{deluxetable*}{ccccc}[t]
\label{tab:spectral_fit}
\tablecaption{Spectral fitting results}
\tablehead{\colhead{Region }& \colhead{Component} & \colhead{Parameter}  & \colhead{Value $\pm1\sigma^c$ error}  & \colhead{Units} }
\startdata
Nucleus             & {\tt cloudy$_l$}   & $\log U_l$   & $-1.4^{+0.3}_{-0.6}$  &      --      \\ 
{\tt cloudy$_h$ + cloudy$_l$ + mytorus} &           & $\log N_{Hl}$  & $22.2^{+0.2}_{-0.7}$  &    $\log(\textrm{atoms cm}^{-2})$        \\ 
wstat/d.o.f. $=1.06$                 &        &  norm$_l$    & $5.4^{+5.1}_{-2.9}$   &    $10^{-14}\times$photons cm$^{-2}$s$^{-1}$keV$^{-1}$         \\ 
{d.o.f.}$^b=199$                   & {\tt cloudy$_h$}    & $\log U_h$   & $0.4^{+0.5}_{-0.6}$    &         --             \\ 
                    &           & $\log N_{Hh}$  & $19.6^{+0.3}_{-0.5}$   &             $\log(\textrm{atoms cm}^{-2})$             \\  
                    &           & norm$_h$     & $7.7^{+7.3}_{-2.9}$    &      $10^{-15}\times$photons cm$^{-2}$s$^{-1}$keV$^{-1}$         \\  
                    & {\tt mytorus}   &  N$_{{H}}$          & $1.7^{+0.3}_{-0.4}$ &   $10^{24}\times$atoms cm$^{-2}$        \\
                    &           &  norm$_{\textrm{c}}$   & $1.5^{+0.4}_{-0.3}$     &    $10^{-2}\times$ photons cm$^{-2}$s$^{-1}$ at 1\,keV            \\
                    &           &  norm$_{\textrm{Fe}}$   & $1.1^{+0.2}_{-0.1}$    &   $10^{-2}\times$ photons cm$^{-2}$s$^{-1}$ at 1\,keV           \\
\hline
Cone Region         &  {\tt cloudy}      & $\log U$   & $-0.5^{+0.1}_{-0.2}$    &      --      \\ 
{\tt xsapec + cloudy  + \tt mytorus$^a$}                     &         & $\log N_H$  & $20.29^{+0.06}_{--}$  &    $\log$(\textrm{cm})$^{-2}$             \\ 
   wstat/d.o.f. $=1.40$       &                & norm       & $4.4^{+2.0}_{-0.9}$   & $10^{-14}\times$photons cm$^{-2}$s$^{-1}$keV$^{-1}$   \\  
   {d.o.f.}$^b=80$           &  {\tt xsapec}    & kT         & $0.94^{+0.04}_{-0.05}$   &        keV    \\ 
                    &           & norm       & $1.5^{+0.1}_{-0.2}$  &  $10^{-5}$norm$_{a}$ \\    \hline    
Cross-Cone Region  &   {\tt xsapec}     & kT         & $0.96^{+0.05}_{-0.05}$ &       keV     \\ 
 {\tt xsapec + cloudy + \tt mytorus$^a$ }                    &   & norm       & $8.6^{+1.0}_{-0.9}$    &  $10^{-6}$ norm$_{a}$ \\ 
               wstat/d.o.f. $=1.15$               & {\tt cloudy}     & $\log U$   & $-0.8^{+0.3}_{-0.7}$  &   -         \\ 
        {d.o.f.}$^b = 50$             &           & $\log N_H$  & $20.0^{+0.6}_{--}$&  $\log$(\textrm{atoms cm}$^{-2}$ )              \\ 
                    &           &  norm    & $7.8^{+0.3}_{-0.7}$  & $10^{-14}\times$photons cm$^{-2}$s$^{-1}$keV$^{-1}$       \\  
                        {\tt xsapec + mytorus$^a$}           &  {\tt xsapec}    & kT         & $1.02^{+0.03}_{-0.06}$   &        keV    \\ 
wstat/d.o.f. $=1.32$    &           & norm    & $1.4^{+0.2}_{-0.1}$   &  $10^{-5}$norm$_{a}$  \\  
 {d.o.f.}$^b$ = $53$   &           &        &   & \\ 
\hline
\enddata
\footnote{{\tt mytorus} accounts for the PSF leakage from the nucleus. Its normalization is fixed to 4\% of the nucleus normalization.  }
\footnote{d.o.f. - degrees of freedom.}
\footnote{{ upper ($+$) and lower ($-$) $1\sigma$ confidence interval.}}
\end{deluxetable*}

\subsubsection{Bi-cones}

Based on the PSF simulations (Sec.\,\ref{subsec:psf}), we expect the bi-cone emission to dominate over the nuclear PSF leakage at energies below $4$\,keV. There is also extended emission present in the $6$--$7$\,keV band. To account for the nuclear PSF contamination in the bi-cone spectra above $4$\,keV, we added a {\tt mytorus} component to each model, with parameters fixed to the best-fit values obtained for the nucleus, freezing the normalization of these components to $\sim4\%$ of the nuclear values. { 4\% is the contribution of the hard PSF to the radii $>1.5^{\prime\prime}$, estimated from the our simulated PSF (see Fig.\,\ref{fig:radial_1}).} This {\tt mytorus} component is associated with instrumental effects and is not the actual characteristics of the diffuse emission. 

 { Initially, we fit the spectra of the two cones separately; however, due to the limited photon statistics, the best-fit model parameters were not well constrained and are consistent with each other within the errors. Therefore, we merged the spectra of the NW and SE cones.} 
We initially allowed for intrinsic absorption ({\tt xsphabs}), but the resulting best-fit column density was only an upper limit, so this component was excluded in the final model. { Single {\tt apec} and single {\tt cloudy} models do not provide satisfactory fit to the data (see Appendix\,\ref{appendix:spectra} for the f-test results).  }
A model consisting of a combination of {\tt apec} and {\tt cloudy} components { leads to the best fit quality, with { {\tt wstat / d.o.f.}}$=1.37$. Models with two {\tt apec} and two {\tt cloudy} components provide poorer fits, with redcued statistic values $1.6$ and $1.8$, respectively. The {\tt apec + cloudy} model is also preferred based on the fit residuals. The {\tt cloudy} model fails to reproduce the spectral shape around $\sim1$\,keV, corresponding to the position of the Ne~IX, Ne~X, and Fe-L line blend 
(see the bottom panel of Fig.\,\ref{fig:spectra_other1}), which is well reproduced by the {\tt apec} model. The purely thermal model, on the other hand, performs worse in describing the shape of the spectrum at energies between $1.5$ and $4.0$\,keV.}

Introducing an additional component to the {\tt apec} and {\tt cloudy} combination does not improve the fit quality. In our resulting fit, the thermal plasma has a best-fit temperature of $kT \sim 1.0$\,keV, and the photoionized gas is characterized by moderate ionization ($\log U \sim 0$) and low column density ($\log N_H \sim 20.8$). Best-fit values of parameters are included in Tab.\,\ref{tab:spectral_fit}.

Consistent with the PSF modeling, the continuum emission above $4$\,keV is well described by a {\tt mytorus} component fixed at the expected $\sim4\%$ of the nuclear normalization. The extended emission detected in one of the cones at $6-7$\,keV energy range is modeled by an iron line included in the {\tt cloudy} model. 

Below $1.5$\,keV the emission is dominated by the {\tt apec} model, while in the $1.5-2.5$\,keV energy range {\tt apec} and {\tt cloudy} emission are comparable, and above $2.5$\,keV {\tt cloudy} dominates.

In $0.3-4.0$\,keV energy range, the de-absorbed luminosity of the thermal plasma emission  is $\sim9.07\times10^{39}$\,erg\,s$^{-1}$, comparable with the de-absorbed luminosity of the { photo}ionized gas emission $\sim1.06\times10^{39}$\,erg\,s$^{-1}$.

\subsection{Cross-cones}
As in the bi-cone region, we account for nuclear PSF leakage by adding a {\tt mytorus} component with best-fit parameter values describing the nuclear emission and normalization fixed at $\sim4\%$ of the nuclear value (based on PSF modeling). 

Spectra extracted from the cross-cone region have significantly lower count statistics. While a single model component can result in a statistically satisfactory fit, it produces correlated residuals. Therefore, we present also a two-component model consisting of both {\tt apec} and {\tt cloudy}. Parameters describing the thermal plasma are well constrained, with a best-fit temperature of $kT\sim0.95$\,keV, such that the {\tt apec} component dominates emission from $0.7-2$\,keV.  { Adding the {\tt cloudy} component does not improve fit statistics significantly, but it removes the correlated fit residuals (see Tab.\,\ref{tab:spectral_fit}).}

\section{Discussion}\label{sec:discussion}

\subsection{The nuclear region}\label{sec:discussion_nuclear}

\begin{figure}[thp!]
    \centering
    \includegraphics[width=0.45\textwidth]{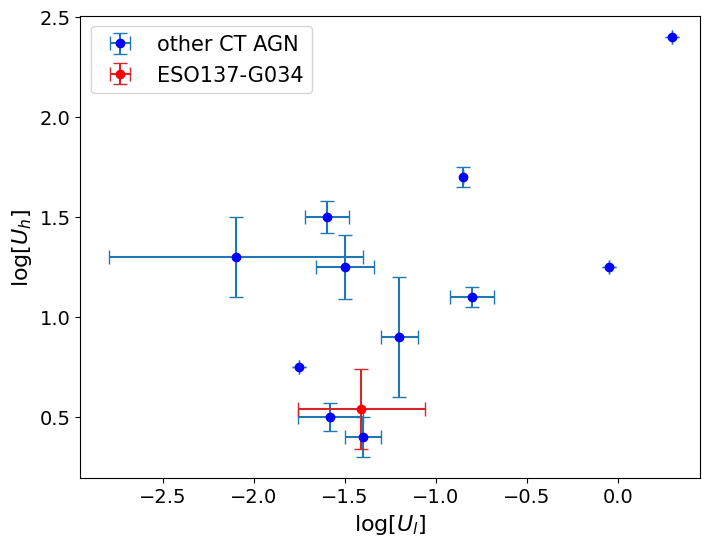}
    \caption{Best-fit ionization of two {\tt cloudy} components fitted to the nucleus ($r<1.5^{\prime\prime}$) spectra, for ESO\,137-G034 and other CT AGN (see Sec.\,\ref{sec:discussion_nuclear} for reference).}
    \label{fig:comparisons}
\end{figure}

As discussed in Sec.\,\ref{sec:nucleus_fit} and shown in Fig.\,\ref{fig:spectra}, the spectrum of the nuclear region below $4$\,keV, { where the direct AGN contribution is likely to be minimal because ESO~137-G034 is a CT AGN}, can be  modeled as emission of at least two { photo}ionized gas phases, a highly ionized, low-column-density component ($\log U_h \sim 0.5$, $\log N_{Hh} \sim 19$) and a low-ionization, high-column-density component ($\log U_l \sim -1.6$, $\log N_{Hl} \sim 22$). In Fig.\,\ref{fig:comparisons}, we compare the best-fit ionization values for ESO\,137-G034 with different CT AGN where a two-component {\tt cloudy} best-fit model has been reported in the literature: ESO\,138-G001 \citep{DeCicco15}, ESO\,428-G001 \citep{Fabbiano18b}, IC\,5063 \citep{Travascio21}, Mrk\,573 \citep{Paggi12}, NGC\,1068 \citep{Kraemer15}, NGC\,3783 \citep{Blustin02}, NGC\,5728 \citep{Trindade23}, and NGC\,7212 \citep{Jones20}. { Fig.\,\ref{fig:comparisons} shows that the ionization parameters we find for ESO 137-G034 are consistent with the relation between the high and low ionization parameters obtained for other CT AGN. These results suggest at least two-phase { photo}ionized medium in the extended emission components of CT AGN, probably arising from a distribution of densities in the ISM of the circum-nuclear region.}

The high-energy part of the nuclear spectrum is described well by the {\tt mytorus}, that models the CT AGN emission where photons from the inner accretion disk and corona are intercepted by the nuclear torus, which gives rise to the reflection spectral component and the fluorescent Fe\,K$\alpha$\, $6.4$\,keV line.

\subsection{The extended spectrally complex bi-cones}

{ As shown in Sec.\,\ref{sec:morphology}, there is a strong extended X-ray emission in ESO~137-G034, displaying the characteristic dual ionization cone morphology commonly observed in Seyfert\,2 galaxies. Extended emission in the bi-cone region is detected at energies up to $5$\,keV and additionally in the $6$--$7$\,keV range, which includes the Fe\,K$\alpha$ emission line. 
In the soft energy band ($<3.0$\.keV) this extended bi-cone emission is detected as far as $2$\,kpc from the nucleus, while at higher energies reaches $\sim800$\,pc (Fig.\,\ref{fig:radial_1}).
The azimuthal profiles of the extended emission show an energy-dependent morphology, with SE cone exhibiting different emission maxima in the $0.3$--$1.5$\,keV and $1.5$--$4.0$\,keV energy ranges (Fig.\,\ref{fig:azimuthal}).

Spectral analysis (Sec.\,\ref{sec:spectra}) showed that the extended emission in the bi-cones is multicomponent, with signatures of both shock-heated thermal and photoionized plasma emission. }

{ The best-fit model describing the soft part of the bi-cone spectrum is {\tt apec} + {\tt cloudy}.  Even though the  photon statistics and CCD's energy resolution limit our abilities to characterize unequivocally the parameters of the emitting gas, based on the spectral fitting results, we can conclude that} the extended emission originates from a mixture of a shock-heated thermal plasma and a { photo-}ionized medium.  As shown in Table\,\ref{tab:fit_other}, a two {\tt cloudy} model or a two {\tt apec} model give a bad fit and therefore are rejected. Thermal plasma emission in particular is needed to account for the spectral shape up to $\sim1$\,keV. Based on the best-fit model, we conclude that hot plasma emission dominates the soft energy band (especially between $0.8$ and $1.5$\,keV), and { photo-}ionized gas { is equally important between $1.5$\,keV and $2.5$\,keV, while it dominates between $2.5$\,keV and $4.0$\,keV (see Fig.\,\ref{fig:spectra})}.

{ From the temperature of the thermal component, }we estimated the velocity of the associated shock as $v\sim800-1000$\,km\,s$^{-1}$ \citep[$v=\sqrt{\frac{16kT}{3\mu}}$, where $k$ is the Boltzmann constant and $\mu$ is the mean molecular mass of a fully ionized gas; e.g.,][]{Fabbiano18b}.  

A single {\tt cloudy} component describing the { photo-ionized plasma } emission in the cone region can be a major simplification. Due to the size of the region, the ionization parameter value may be different at different distances from the nucleus. { This effect, however, depends on the ISM density profile and is mitigated if the density follows \(\propto r^{-2}\), a slope often found in AGN winds \citep[][]{Bianchi06}.}

 To spatially disentangle { thermal and photo-ionized emission regions}, we conducted a { spatial} analysis of the energy-dependent emission morphology.

\subsubsection{The [O III]/ X-ray ratio diagnostic}

\begin{figure*}[t!]
    \centering     
    \includegraphics[width=0.49\textwidth]{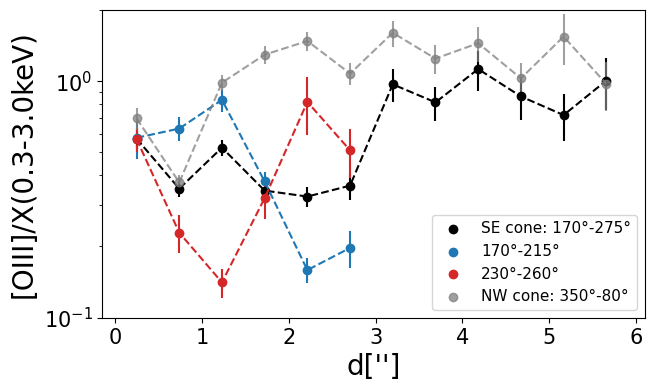}
    \includegraphics[width=0.49\textwidth]{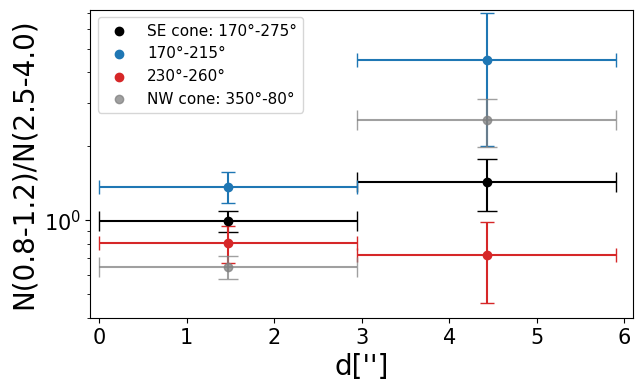} 
\caption{ {Left panel: O[III]/X($0.3-3.0$\,keV) flux ratio  for the SE cone (black points), NW cone (grey points) and two regions within SE (red and blue, corresponding to cones marked in red and blue in Fig.\,\ref{fig:azimuthal}. Right panel: The X-ray hardness ratio for SE cone (black points), NW cone (grey points) and two regions within SE, red and blue (see Fig.\,\ref{fig:OIII}). }}
\label{fig:OIII}
\end{figure*}
{ Our first approach is to examine} the radial profile of the [O~III] to soft X-ray flux ratio { in different spatial regions. This ratio is a proxy of the ionization parameter for a photo-ionized gas, and  is expected to have an approximately power-law shape \citep{Bianchi06}. In the case of photoionized AGN winds, the [O III]/X-ray ratio is expected to be constant, given the $\propto r^{-2}$ radial dependence of both the ionizing flux and the gas density. This radially constant [O III]/X-ray ratio has been observed in NGC~4151 \citep{Wang09,Wang11b}.}

We investigated the relation between [O~III] \citep[][]{Ferruit00} and the $0.3-3.0$\,keV X-ray morphology in the ESO\,137-G034 bi-cone region. Following the approach of \cite{Paggi12}, we used a conversion factor of $4.6\times10^{-12}$, based on the spectral modeling of the cone emission, to convert cts s$^{-1}$ to {erg cm$^{-2}$ s$^{-1}$.} 

{ We used the entire SE and NW cone regions, and the subregions, marked in blue and red, which were  chosen based on the positions of the double peaks in the azimuthal profile of the SE cone emission, appearing in the soft and medium energy ranges respectively ({ see Fig.\,\ref{fig:azimuthal} in }Sec.\,\ref{sec:azimuthal}). These subregions are defined by the} position angles: $170^{\circ}-215^{\circ}$ - blue part of the SE cone, $230^{\circ}-260^{\circ}$ - red part of the SE cone.

The [O~III]/X-ray profiles extracted for the full NW and SE cone regions and the sub-regions { are plotted in the left panel of  Fig.\,\ref{fig:OIII}.  In the SE cone, the  full cone profile (in black) has a lower [O~III]/X-ray ratio in the inner $3^{\prime\prime}$ ($\sim600$\,pc). A similar [O~III] decrement is only observed in the inner $\sim1^{\prime\prime}$ ($\sim200$\,pc) of the NW cone (plotted in grey). This lower ratio suggest the presence of additional non-photoionized (i.e. thermal)  X-ray emission \citep[see ][]{Wang09}. At larger radii ($>3^{\prime\prime}$), the [O~III]/X-ray ratio for the SE cone (black points) and NE cone (grey points) are constant, consistent with the expectations from photoionized wind emission. However, the situation appears more complex if we examine the red and blue subregions of the SE cone. The [O~III]/X-ray profile behaves different, with ratio increasing between  $1^{\prime\prime}$ and $3^{\prime\prime}$ for the red region and decreasing for the blue region, suggesting a more complicated spatial distribution of different parameters describing the emission, e.g., relative contributions of shock-heated thermal and photoionized plasma components, plasma temperature or absorption.  }

\subsubsection{{ X-ray spectral band ratio imaging}}

\begin{figure}[t!]
    \centering  
    \includegraphics[width=0.49\textwidth]{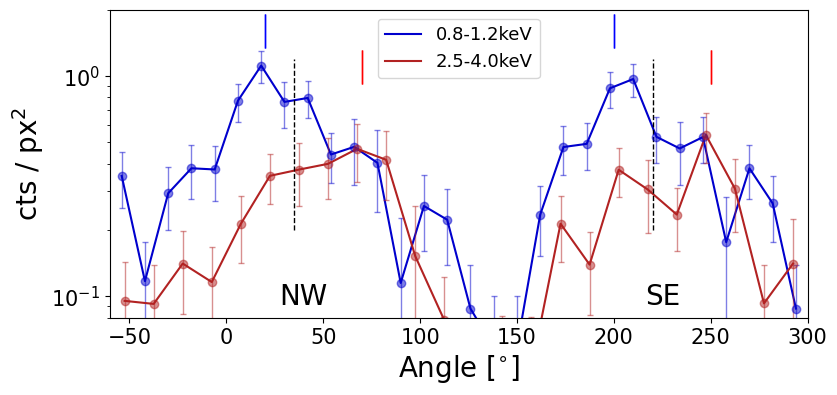}
    \includegraphics[width=0.49\textwidth]{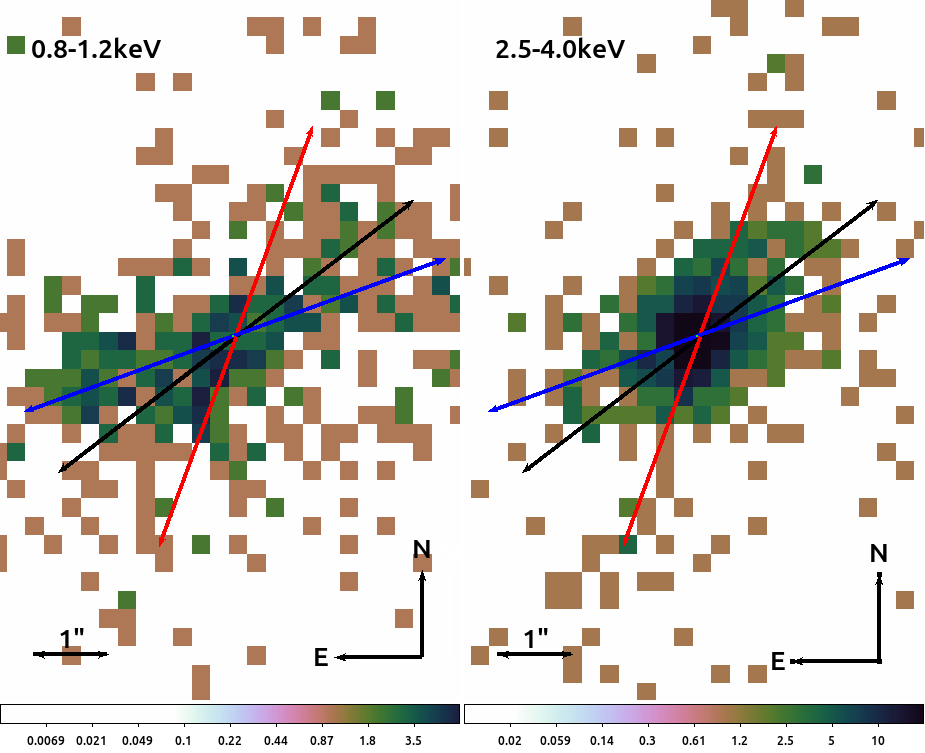}
\caption{ { Lower panel: images of ESO~137-G034 in energy bands energy bands dominated by {\tt apec} ($0.8-1.2$\,keV , left panel) and {\tt cloudy} ($2.5-4.0$\,keV, right panel). Images were binned to $0.5$ of the original {\it Chandra} pixel size. Upper panel: azimuthal profiles of the extended emission between $1.5^{\prime\prime}$ and $9.0^{\prime\prime}$ from the nucleus for corresponding energy ran in bi-cone spectra. The red and blue vertical lines mark maxima of azimuthal profiles. Their angular position are marked in lower panels. The black lines mark the positions of the maximum for $0.3-7.0$\,keV energy range.  }}
\label{fig:HR}
\end{figure}

{ 
The ACIS CCD resolution, coupled with the exquisite Chandra imaging capabilities  can be used to explore spectrally different features in the extended X-ray emission \citep{Wang11a,Paggi12,Fabbiano18b}. We follow this approach here, using the spectral fit results to define energy bands dominated by either thermal or photoionized emission. 

Using the best-fit model for spectrum of the combined cone emission (Fig.\,\ref{fig:spectra}) as a guide, we select spectral bands where the two different emission models dominate. For the thermal emission of shock-heated plasma, we choose the $0.8-1.2$\,keV band, where the thermal contribution exceeds the photoionization contribution by a factor of $\sim10$. This band includes the Ne~IX, Ne~X, and Fe-L emission \citep[]{Wang11a}. For the photoionized contribution we use the $2.5-4.0$\,keV band, where the cloudy contribution is dominant. }

{ The right panel of Fig.\,\ref{fig:OIII} presents the radial profile of X-ray hardness ratio, defined as N($0.8-1.2$)/N($2.5-4.0$), where N($0.8-1.2$) and N($2.5-4.0$) are a number of counts in $0.8-1.2$\,keV and $2.5-4.0$\,keV energy ranges respectively. Profiles are plotted for full SE (black) and NW (grey) cones and SE cone subregions (blue and red, see bottom-right panel of Fig.\,\ref{fig:azimuthal}). In the inner part of the source ($<3^{\prime\prime}\sim600$\,pc) the hardness ratio is higher in the SE cone (black points) than in NW cone (grey points). The excess originates from the SE subregion marked in blue, suggesting a higher contribution of shock heated thermal plasma emission in that part of the source.  }

{ To further examine the spatial separation of the thermal and photoionized emission, based on our spectral modeling results, we extracted azimuthal profiles (upper panel of Fig.\,\ref{fig:HR}) for the soft and medium energy bands, as previously done in Sec.\,\ref{sec:azimuthal}. This time, following the spectral fitting results, we adjusted the energy ranges to $0.8$--$1.2$\,keV and $2.5$--$4.0$\,keV. This revealed a clear offset of the soft (blue) emission maxima, located at $\sim20^{\circ}$ and $\sim200^{\circ}$, from the maxima determined using the full energy range ($0.3$--$7.0$\,keV), located at $\sim35^{\circ}$ and $\sim215^{\circ}$ (vertical dashed lines), not only in the SE cone but also in the NW cone. The azimuthal profile for the $2.5$--$4.0$\,keV energy band (red) more closely resembles that of the full energy band, but shows hints of a maxima offset in the opposite direction to the soft-band emission at $\sim70^{\circ}$ and $\sim250^{\circ}$.   This energy dependence in the direction of the emission extension is also visible in the images. In the lower panel of Fig.\,\ref{fig:HR} we present images of the ESO~137-G034 X-ray emission in $0.8-1.2$\,keV (left) and $2.5-4.0$\,keV (right) energy ranges binned with the $0.5$ original {\it Chandra} pixel size, with lines marking the positions of the maxima in the azimuthal profiles.}

\subsubsection{Radio jet emission}

\begin{figure}[t!]
    \centering  
    \includegraphics[width=0.49\textwidth]{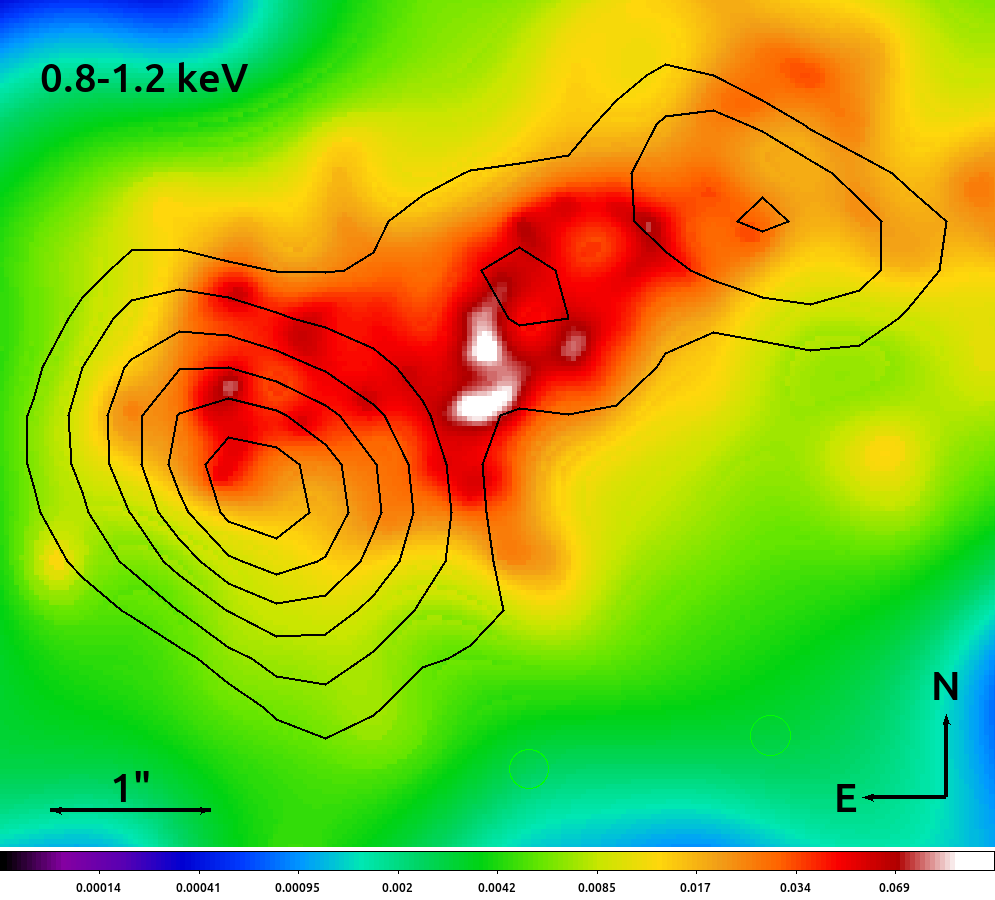}
\caption{ { Adaptively smoothed X-ray image of ESO~137-G034 in the $0.8$--$1.2$\,keV energy band. The $3$\,cm radio jet emission \citep{Morganti99} is marked with black contours. }}
\label{fig:radio}
\end{figure}

{ In Fig.\,\ref{fig:radio} we compare the $0.8-1.2$\,keV, thermal emission dominated image, with the position of the $3$\,cm radio jet of ESO~137-G034 from the Australia Telescope Compact Array \citep[ATCA, half-power beam width = $1.271^{\prime\prime}$][]{Morganti99}. Contour levels start at the $3\sigma$ level, where $\sigma$ is the standard deviation of the background-dominated region, and increase by a factor of $\sqrt{2}$. The radio jet aligns well with the structure of the extended X-ray emission. It is particularly pronounced in the SE cone region, extending up to $\sim3^{\prime\prime}$ ($\sim600$\,pc) from the nucleus, in the area associated with the [O~III]/X-ray ratio decrement (Fig.\,\ref{fig:OIII}) and the soft emission excess seen in the X-ray hardness ratio map (Fig.\,\ref{fig:HR}). This suggests that the thermal emission component may arise from the shocks induced by the propagation of the radio jet, further reinforcing the conclusion of a strong shock presence in the SE cone. The presence of strong shocks in this regions is also confirmed by the optical {\it HST} ionization map (Król et al., submitted). The difference can be also partially associated with the extinction of the [O~III] emission due to the dust. }

The stronger shock presence in the SE cone { would be } consistent with findings of \cite{Zhang24L}, who studied regions roughly corresponding to the inner part SE and NW cones defined in our analysis through polycyclic aromatic hydrocarbon (PAH) diagnostics. The SE cone appears to have a lower fraction of small and neutral grains compared to the NW cone and the nucleus. Since small, ionized PAHs are most vulnerable to destruction, this indicates differing environmental conditions. {Moreover,  \cite{Zhang24} report strong shock signatures in ESO~137-G034 based on broad emission line profiles, with line widths containing the central $80$\% of the flux,  W80 $\gtrsim 500$\,km s$^{-1}$) for  [Ar~II,III], [Ne~II,III,V] and [S~IV].  

Assuming that shock-heated thermal plasma emission dominates the X-ray emission in the vicinity of the SE radio jet,} we extracted the spectra from the part of the SE cone corresponding to the radio-jet position and the dip in the [OIII]/X-ray flux ratio, between $1.5^{\prime\prime}$ and $3.5^{\prime\prime}$ {to study its physical parameters}. 

{  As in Sec.\,\ref{sec:spectra}, we fitted the extracted spectrum with a {\tt apec+cloudy} model. kT and $\log U$ were frozen at the best-fit values obtained in Sec.\,\ref{sec:spectra}, and only model normalizations were allowed to vary. We obtained a fit characterized by $\chi^2=1.27$, with the best-fit  {\tt apec} norm$ = (1.28\pm0.23)\times10^{-5}$\,norm$_a$. } Assuming that the emitting region is cone-shaped, with the main axes on the plane of the sky and is filled with ionized hydrogen, we derive the gas density $n_H\sim0.27f^{-0.5}$\,cm$^{-3}$, leading to an emitting gas mass of $\sim8\times10^{5}f^{0.5}$\,M$_{\odot}$, where $f$ marks the fraction of the thermal emitting matter within the total region volume ($f<1$). The pressure $p=n_HkT$ is estimated as $\sim4\times10^{-10}f^{-0.5}$\,dyn\,cm$^{-2}$, and the internal energy $\epsilon_{int}=1.5pV\sim2\times10^{53}f^{-0.5}$\,erg.

{ Assuming the emitting plasma constitutes a radio jet cocoon, and that all the kinematic energy transported by the jet during its lifetime is converted at the jet head (terminal shock) into the cocoon’s internal pressure, we can estimate the jet power, $L_j$, from $\epsilon_{\rm int} = 1.5pV = L_j t$, where $t$ is the jet lifetime \citep[][]{Stawarz08}. Assuming a jet expansion velocity of $0.3c$, a value typical for Compact Symmetric Objects (CSOs) with sizes $\lesssim500$\,pc \citep[as the jet of ESO~137-G034, ][]{An12} and assuming $\epsilon_{\rm int}\sim2\times10^{53}f^{-0.5}$\,erg from the X-ray spectral fitting, we estimate a lower limit to the jet power of $L_j \sim 10^{42}\sqrt{f}$\,erg\,s$^{-1}$, far below the $\sim10^{44}$\,erg\,s$^{-1}$ threshold required for the jet to expand beyond its host galaxy \citep{Mukherjee16,Mukherjee17}. The inferred jet power is comparable to that modeled for the weakest CSOs \citep{Krol24,Wojtowicz20}. 

Furthermore, we estimate the jet production efficiency, $\eta_j$, as:
\begin{equation}
    \log\eta_j = \log(L_j \eta^{-1} c^{-2}) \simeq \log \left( 0.1 L_j L_{\rm bol}^{-1} \right) \sim -2.72,
\end{equation}
assuming an efficiency $\eta = 0.1$ \citep{King08} and the disk bolometric luminosity from the approximation:
\begin{equation}
    \log L_{\rm bol} = 0.0378 (\log L_{X})^2 - 2 \log (\log L_{X}) + 60.5,
\end{equation}
where $L_{X}$ is the intrinsic X-ray luminosity in the $2$--$10$\,keV energy range \citep{Ricci17,Ichikawa17}. The low jet production efficiency, $\log\eta_j \sim -2.72$, is comparable to that inferred for CSOs \citep{Wojtowicz20}. }

\subsection{The emission mechanism of the extended cross-cone region}

 Extended emission of ESO\,137-G034 is found also in the cross-cone direction (Sec.\,\ref{sec:morphology}), up to $\sim3$\,keV. Extended emission in the cross-cone direction { has been} found in multiple CT AGN \citep[see, e.g.,][]{Fabbiano18b,Jones21}. It has been explained either in terms of nuclear emission { escaping} through a porous torus \citep{Nenkova02} or as a result of the jet interaction with a dense and patchy ISM \citep{Mukherjee18,Fabbiano22}.

{  The cross-cone spectrum is adequately fitted with a single thermal component with kT$\sim1$\,keV (Table\,\ref{tab:spectral_fit}). It can also be fitted with a thermal + photoionization model similar to that of the cone emission, but the small $\chi^2\sim0.55$ suggests overfitting. These spectral results cannot discriminate between the porous torus and jet interaction options, given the compatibility with the cone spectrum and the much more limited statistics. However, the presence of an embedded radio jet and the shock/turbulent velocities inferred in this paper and in \cite{Zhang24} from MIRI [N II] observations, would favor the jet interaction possibility. }

\section{Summary}\label{sec:summary}

In this paper, we { report the results of our analysis of the} deep {\chandra} ACIS observations of ESO\,137-G034. Our main findings are summarized as follows:
\begin{itemize}

\item { Extended { X-ray emission is detected with a biconical morphology up to $5$\,keV and in the $6$--$7$\,keV energy range. The cross-cone regions show extended emission up to $3$\,keV. In the soft energy band ($< 3$\,keV) the emission extends out to $\sim2$\,kpc, while at higher energies can be traced out to $\sim800$\,pc in the bicones only}.}

\item The nuclear ($r<1.5^{\prime\prime}\sim 300$\,pc) emission below $4$\,keV originates from { photo}ionized gas and can be modeled by at least two gas phases: { one highly ionized with low column density, dominating the emission in the $1.5-4.0$\,keV energy range, and one low-ionization phase with higher column density, dominating the emission below $1.5$\,keV.} Above $4$\,keV, reflection from a toroidal structure (modeled with {\tt mytorus}) dominates. The derived ionization parameters are consistent with those obtained for other CT AGN.

\item Spectral modeling shows that the bi-cone emission is a mixture of both photoionized gas (radiative feedback) and shock-heated plasma (kinematic feedback), with best-fit parameters of $\log U\sim-0.6$ and $kT\sim0.9$\,keV. Both components have comparable luminosities of $\sim9\times10^{39}$\,erg\,s$^{-1}$. { The thermal plasma component is required to reproduce the spectral shape around $\sim1$\,keV, dominating the $0.8$--$1.2$\,keV band by a factor of $\sim10$ over the photoionized contribution. In contrast, photoionization dominates the $2.5$--$4.0$\,keV band.}

\item { The extended emission morphology is energy-dependent, with different azimuthal profile maxima for the $0.8$--$1.2$\,keV and $2.5$--$4.0$\,keV bands. This indicates that distinct physical mechanisms dominate in different regions of the source.}

\item { The radial [O~III]/X-ray flux ratio suggest that the photoionized gas is present in both cones, dominating in the NW cone at $r>400$\,pc and in the SE cone at $r>800$\,pc. At these radii, the profile is consistent with a $\rho\propto r^{-2}$ density law, implying radiative feedback dominance far from the nucleus. Localized X-ray excesses at smaller radii indicate additional thermal emission.}

\item { In the SE cone, an X-ray excess coincides with the radio jet. Hardness ratio maps reveal a counts excess in the $0.8$--$1.2$\,keV band (Ne and Fe-L lines) aligned with the jet, supporting a shock origin. In this region, we estimate a lower limit to the thermal plasma internal energy of $\sim2\times10^{53}$\,erg and a density of $\gtrsim0.27$\,cm$^{-3}$, corresponding to a mass of $\sim8\times10^5$\,M$_\odot$. These results suggest kinematic feedback plays a major role within $\sim800$\,pc, especially in the SE cone.  Assuming this plasma forms a radio jet cocoon, we infer a jet power of $L_j\sim10^{42}\sqrt{f}$\,erg\,s$^{-1}$ and a low jet production efficiency of $\log\eta_j\sim-2.72$, values comparable to those found for the weakest Compact Symmetric Objects.}
\end{itemize}

\software{CIAO v.4.16 \citep{Fruscione06},
Sherpa \citep{Siemiginowska24, Doe07,Freeman01}, MARX \citep{Davis12}, astropy  \citep{Astropy13},  Cloudy \citep{Ferland13}, }

\vspace{5mm}
\facilities{CXC, ATCA, HST}

\begin{acknowledgments}
We thank R. Morganti for sharing the ATCA 3 cm data used in this work. This work was supported by the Chandra Guest Observer program grant GO2-23075X, and partially by NASA contract NAS8-03060 (CXC). It was performed in part at Aspen Center for Physics, which is  supported by National Science Foundation grant PHY-2210452.
\end{acknowledgments}

\clearpage
\appendix

\restartappendixnumbering
\section{Alternative spectral models}\label{appendix:spectra}
\begin{figure*}[thp!]
    \centering   
    \includegraphics[width=0.45\textwidth]{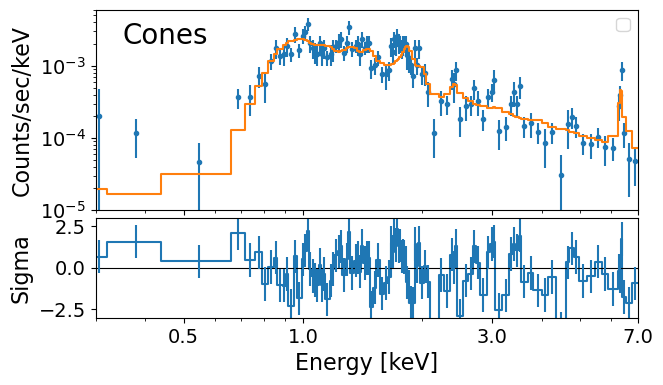}
    \includegraphics[width=0.45\textwidth]{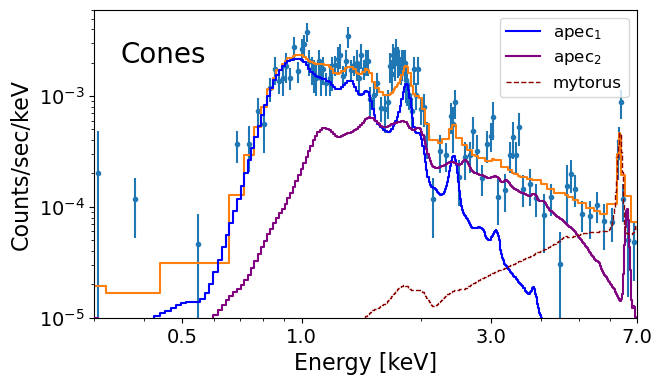}
    \includegraphics[width=0.45\textwidth]{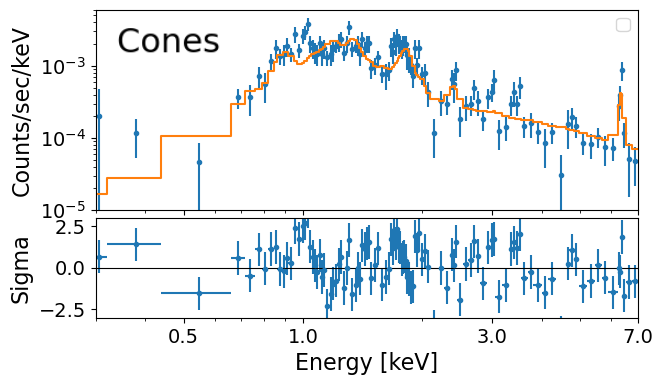}
    \includegraphics[width=0.45\textwidth]{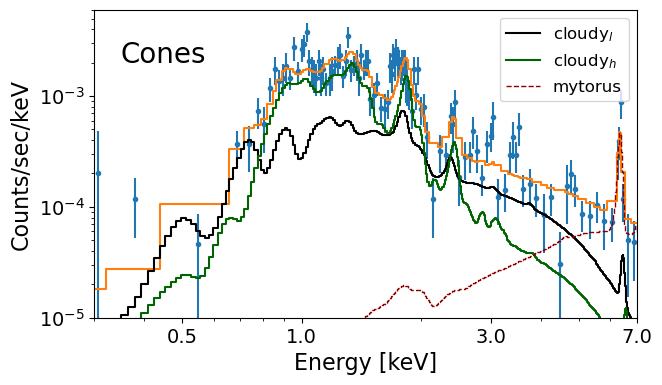}
    \includegraphics[width=0.45\textwidth]{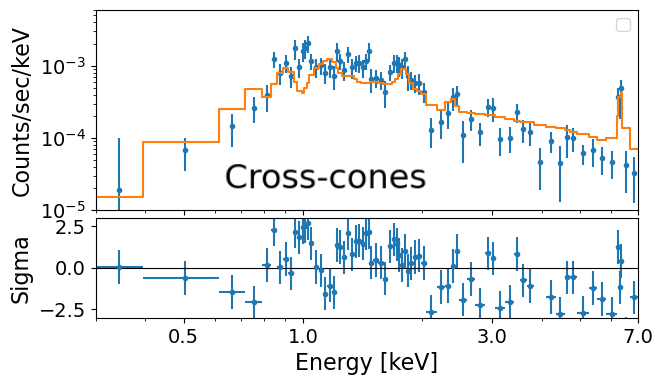}
    \includegraphics[width=0.45\textwidth]{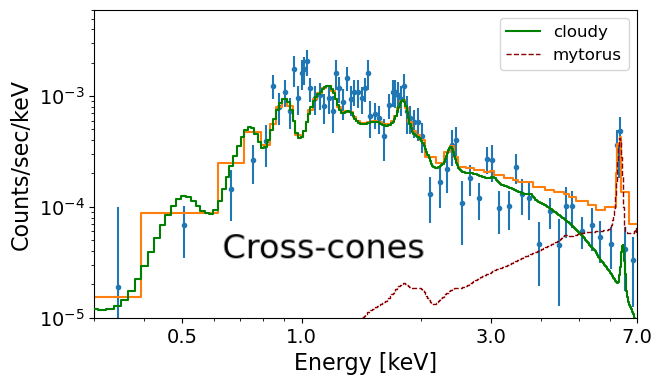}
    \caption{Spectrum of the bi-cone (first and second row) and cross-cone (bottom row) regions. The left column shows the best-fit of rejected models and their residua, while the right column presents the model components. }
    \label{fig:spectra_other1}
\end{figure*}

{ In Tab.\,\ref{tab:fit_other}, we list the best-fit parameters and fit statistics of the rejected spectral models describing the emission from the nucleus, bi-cone, and cross-cone regions. A description of the model components and their parameters can be found in Sec.\,\ref{sec:models_components}. In Fig.\,\ref{fig:spectra_other1}, we show the regions spectra along with rejected best-fit models, their residuals, and individual model components.}

{ Single-component models describing the emission below $4$\,keV can be rejected for both the nucleus and bi-cone spectra with high statistical significance, based on the F-test comparison with the best-fit two-component models. For the nucleus, the significance is $\sim10^{-25}$ and $\sim10^{-6}$ for the comparison of single {\tt apec} and {\tt cloudy} models, respectively, with the two {\tt cloudy} models. For the bi-cones, the significance is $\sim10^{-6}$ for the {\tt apec} model and $\sim10^{-12}$ for the {\tt cloudy} model, both with respect to the two-component {\tt cloudy + apec} model. }

{ The two {\tt apec} model was rejected for the nucleus based on the reduced { wstat} value and the presence of correlated residuals between $1$ and $3$\,keV. The {\tt apec + cloudy} requires non realistically high intrinsic {\tt apec} luminosity.   The bi-cone spectra cannot be described by either two {\tt cloudy} or two {\tt apec} components. The two {\tt cloudy} components fail to reproduce the spectral shape around $\sim1$\,keV, whereas the two {\tt apec} components show residuals at $\sim1.5$--$3.0$\,keV (see the second and third rows of Fig\,\ref{fig:spectra_other1}). The fit statistics are also better for the {\tt apec + cloudy} model. The fit statistics of the single component {\tt cloudy} model for the cross-cone region is significantly worse ({wstat} / d.o.f.$=1.98$), than in the case of single {\tt apec} model. { Parameters definitions are given in Sec.~\ref{sec:models_components}}.}

\begin{deluxetable*}{ccccc}[b]
\label{tab:fit_other}
\tablecaption{Rejected spectral models.}
\tablehead{\colhead{Region }& \colhead{Component} & \colhead{Parameter}  & \colhead{Value $\pm1\sigma$ errors$^b$}  & \colhead{Units} }
\startdata
Cone               &  {\tt xsapec$_1$}    & kT$_1$         & $0.8^{+0.1}_{-0.1}$   &        keV    \\ 
{\tt xsapec$_1$ +  xsapec$_2$ + mytorus$^{a}$} &           & norm$_1$       & $1.7^{+0.1}_{-0.2}$   &  $10^{-5}$norm$_{a}$  \\ 
   wstat/d.o.f. $=1.68$    &      {\tt xsapec$_2$}    & kT$_2$         & $4.0^{+3.2}_{-1.2}$   &        keV    \\ 
     d.o.f.$^b = 79$               &           & norm$_2$       & $1.2^{+0.2}_{-0.2}$  & $10^{-5}$norm$_{a}$  \\    
{ \tt cloudy$_h$ + cloudy$_l$ + mytorus$^{a}$}               & {\tt cloudy$_l$}   & $\log U_l$   & $-0.4^{+0.1}_{-0.4}$  &      --      \\ 
   wstat/d.o.f. $=1.76$             &           & $\log N_{Hl}$  & $20.0^{+2.1}_{-0.6}$  &    $\log$(\textrm{atoms cm}$^{-2})$        \\ 
    d.o.f.$^b = 77$                   &           &  norm$_l$    & $5.1^{+7.3}_{-4.7}$   &    $10^{-14}\times$photons cm$^{-2}$s$^{-1}$keV$^{-1}$          \\ 
                    & {\tt cloudy$_h$}    & $\log U_h$   & $1.6^{+0.1}_{-0.1}$    &         --             \\ 
                    &           & $\log N_{Hh}$  & $19.6^{+1.1}_{--}$   &             $\log$(\textrm{atoms cm}$^{-2}$ )             \\  
                    &           & norm$_h$     & $4.4^{+3.7}_{-1.2}$    &      $10^{-16}\times$photons cm$^{-2}$s$^{-1}$ keV$^{-1}$        \\  
\hline
Cross-Cone             & {\tt cloudy}   & $\log U$   & $-0.19^{+0.09}_{-0.03}$  &      --      \\ 
 { \tt cloudy+ mytorus$^{a}$}                    &           & $\log N_H$  & $20.3^{+0.3}_{-0.2}$  &    $\log$(\textrm{atoms cm}$^{-2})$        \\ 
     Reduced { wstat} / d.o.f.$=1.98 $                 &           &  norm$_l$    & $2.1^{+1.0}_{-0.8}$   &    $10^{-14}\times$photons cm$^{-2}$s$^{-1}$keV$^{-1}$          \\ 
     d.o.f.$^b = 52$    & & & & \\ 
\hline     
\enddata
\footnote{{\tt mytorus} accounts for the PSF leakage from the nucleus. Its normalization is fixed to 4\% of the nucleus normalization.  }
\footnote{{ upper ($+$) and lower ($-$) $1\sigma$ confidence interval.}}
\end{deluxetable*} 

\bibliography{sample631}{}
\bibliographystyle{aasjournal}

\end{document}